\documentclass{article}
\usepackage{amssymb}
\usepackage{amsfonts}
\usepackage{amsmath}
\usepackage{graphicx}

\begin{document}

\title{Characterizing planetary orbits and trajectories of light in the
Reissner-Nordstr\"{o}m metric}
\author{F.T. Hioe* \and Department of Physics, St. John Fisher College,
Rochester, NY 14618}
\maketitle

\begin{abstract}
Exact analytic expressions for planetary orbits and light trajectories in
the Reissner-Nordstr\"{o}m geometry are presented. They are characterized in
a map specified by three dimensionless parameters for the planetary orbits,
while two dimensionless parameters are required to map the trajectories of
light. Notable differences with the corresponding orbits and trajectories in
the Schwarzschild geometry are indicated. In particular, when the energy and
angular momentum of the planet are fixed, the precession angle of the orbit
decreases as the net electric charge of the massive star or black hole
increases. A similar result also holds for the deflection angle of a light
ray.

\end{abstract}

\section{Introduction}

It is well known that besides the Schwarzschild spherically symmetric
solution of Einstein's equation for the vacuum, there is the Reissner-Nordstr%
\"{o}m (R-N) spherically symmetric solution of the coupled equations of
Einstein and Maxwell [See e.g.1,2]. The R-N geometry applies to a massive
object or a black hole with mass $M$ and electric charge $Q$. Since we have
not observed any large macroscopic body in the universe that possesses a net
charge, the consideration of a charged massive object or black hole would
appear to be unrealistic. Nevertheless, the study of the R-N solution is
useful to our understanding of the nature of space and time. At the very
least, one would like to know what the most notable effect of the presence
of a net electric charge on a massive object is on the trajectory of an
electrically neutral planet (which we shall call a particle) or a light ray
outside the massive object. We shall refer to the massive object as a black
hole even though most of the results that we present in this paper apply
equally if it is simply a massive star. With the assumption that the charged
black hole is centered at the origin of the coordinates, we shall present
analytic expressions for all trajectories of particles and light in the
polar coordinates $(r,\phi )$ in the equatorial plane $\theta =\pi /2$.
Because the time coordinate has been eliminated and is not present in our
expressions, we do not discuss in this paper the many delicate questions
about time when the particles or photons may cross the event horizon. We
concentrate on the method that we have taken [3, 4] of putting all possible
trajectories onto a universal map characterized by three dimensionless
parameters and divided into regions with clearly defined analytic
boundaries. On any given point of this map, the corresponding trajectory can
be expressed in a simple analytic expression in terms of the Jacobian
elliptic functions [5] with a predictable behavior. We shall also highlight
the principal differences the electric charge on the black hole makes on the
trajectories and how our analytic expressions reveal these differences.

Analytic expressions have been used before to describe the geodesics in
Schwarzschild and in Reissner-Nordstr\"{o}m metrics, mainly in terms of
Weierstrass elliptic functions [5]. The Schwarzschild metric was treated
using these functions in the early work of Hagihara [6] and Whittaker [7]
while the R-N metric was analyzed in more recent papers for particles [8-11]
and for light rays [12, 13] [see also other references therein]. As will
become clear from the remainder of this Introduction and this paper, the
ways these authors analyzed their analytic expressions and classified their
trajectories are quite different from the method given in this paper and in
our earlier papers [3, 4]. Our approach and treatment may be thought of as
giving an alternative and useful perspective, in addition to some specific
results that were not given previously. We should also mention that one of
the earliest (but very brief) analytic works that made use of the Jacobian
elliptic functions for the Schwarzschild metric is that of Forsyth [14].

We now discuss in more detail the approach that we use in this paper.
Instead of the common practice of using the total energy, angular momentum,
and generalized geometric eccentricity (which Chandrasekhar called $E$, $L$
and $e$ respectively in ref.2) for characterizing different trajectories in
the Schwarzschild geometry, it was suggested in refs.3 and 4 that it is more
convenient to put all possible trajectories of particles on a map specified
by two dimensionless parameters. We follow the same procedure here but add
an additional dimensionless parameter for the electric charge for the R-N
geometry. We do not use units for which the universal gravitation constant $%
G $ and the speed of light $c$ are set equal to one, and, as in our previous
work, we ignore the effect of gravitational radiation.

For particles in the Schwarzschild geometry, the two dimensionless
parameters that we choose to represent the coordinates of the map are $e^{2}$
and $s^{2}$ which we called the energy and field parameters respectively
that are defined by

\begin{equation}
e^{2}\equiv 1+\frac{h^{2}c^{2}(\kappa ^{2}-1)}{(GM)^{2}}=1+\frac{\kappa
^{2}-1}{s^{2}},
\end{equation}

and

\begin{equation}
s^{2}\equiv \left( \frac{GM}{hc}\right) ^{2},
\end{equation}

where $\kappa $ is the total energy per unit rest energy of the particle,
and $h$ is the angular momentum per unit rest mass of the particle. The
coordinates of this universal map for all possible particle trajectories are
$-\infty <e^{2}<+\infty $ and $0\leq s^{2}\leq +\infty $ [15]. One may of
course use the coordinates $(\kappa ,s)$ with $0\leq \kappa \leq +\infty $
and $0\leq s\leq +\infty $ for the map [see ref.4 for some description of
this]. Unlike $\kappa $ and $s$ that represent two independent physical
quantities, namely, energy and angular momentum, $e^{2}$ is a (special)
combination of these two quantities. The advantage of using $(e^{2},s^{2})$
is that $e^{2}$ not only is a convenient combination of the total energy and
angular momentum of the particle for all possible trajectories, but also
that for small values of $s^{2}$, $e$ represents the true geometrical
eccentricity of the Newtonian orbits for the entire range of $+\infty \geq
e\geq 0$, and it alone, without $s$, can be used to characterize the
Newtonian orbit of the particle. We should emphasize that in general our
parameter $e$ is not the same as the generalized geometrical eccentricity
used by Darwin [16], Chandrasekhar [2] and many researchers (also denoted by
$e$ but its definition is different and its value can be complex the way
they defined it). In the Newtonian limit, $c^{2}(\kappa ^{2}-1)$ in eq.(1)
becomes $2E_{0}/m_{0}$, where $E_{0}$ is the sum of the kinetic and
potential energies of the particle, $m_{0}$ is the rest mass of the
particle, and $e$ becomes

\begin{equation}
e\simeq \sqrt{1+\frac{2E_{0}h^{2}}{m_{0}(GM)^{2}}}
\end{equation}

and is the geometrical eccentricity of the particle orbit. We should also
emphasize that we generally use $e^{2}$ instead of $e$ because $e^{2}$ can
be negative. Indeed the negative values of $e^{2}$ are indicative of a
non-classical general relativistic region that allows elliptic orbits with a
non-Newtonian eccentricity [17]. The radial distance $r$ of the particle
from the black hole is expressed as

\begin{equation}
q=\frac{r}{\alpha },
\end{equation}

in units of the Schwarzschild radius $\alpha $ defined by

\begin{equation}
\alpha \equiv \frac{2GM}{c^{2}}.
\end{equation}

The parameter space $(e^{2},s^{2})$ is divided into two regions which we
called Regions I and II [3,4]. Region I has bound, unbound, and terminating
orbits, and Region II has terminating orbits only.

For particles in the R-N geometry, we need a third dimensionless parameter
to measure the charge on the black hole. Noting that $GQ^{2}/(4\pi \epsilon
_{0}c^{4})$, where $\epsilon _{0}$ is the permittivity of free space, has
the dimension of $(length)^{2}$, we define the dimensionless parameter

\begin{equation}
\beta ^{2}\equiv \frac{GQ^{2}}{4\pi \epsilon _{0}c^{4}\alpha ^{2}},
\end{equation}

to be a measure of the electric charge on the black hole. Instead of using a
parameter space with three coordinates $(e^{2},s^{2},\beta ^{2})$ that are
difficult to view, we continue to use a parameter space with two coordinates
$(e^{2},s^{2})$ for a given value of $\beta ^{2}$, and show that the
parameter space can again be divided into two regions but the boundary
depends on the value of $\beta ^{2}$. We will see that it suffices for us to
show the boundary separating the two regions for only two extreme cases for $%
\beta ^{2}=0$ and $\beta ^{2}=1/4$ from which one can see where the
approximate boundaries are for the intermediate values of $\beta ^{2}$.

For the trajectories of light in the Schwarzschild geometry, it was shown in
refs.3 and 18, following the suggestion of Martin [19], that they can be
characterized by a single dimensionless parameter (denoted by $U_{1}$) which
is related to the ratio $\kappa /h$ of the energy and angular momentum of
light. For the R-N geometry, we use the parameter space $(U_{1},\beta ^{2})$
and obtain the specific boundaries that divide the space into three regions.

\section{Particle Trajectories in the R-N Geometry}

We consider the Reissner-Nordstr\"{o}m (R-N) geometry, i.e. the static
spherically symmetric gravitational field in the space surrounding a massive
spherical object such as a star or a black hole of mass $M$ carrying a net
electric charge $Q$. The R-N metric for the spacetime outside the black hole
in the spherical coordinates $r,\theta ,\phi $ is [1,2]

\begin{equation}
dl^{2}=c^{2}\left( 1-\frac{\alpha }{r}+\frac{Q_{\ast }^{2}}{r^{2}}\right)
dt^{2}-\left( 1-\frac{\alpha }{r}+\frac{Q_{\ast }^{2}}{r^{2}}\right)
^{-1}dr^{2}-r^{2}d\theta ^{2}-r^{2}\sin ^{2}\theta d\phi ^{2}
\end{equation}

where $\alpha $ is the Schwarzschild radius defined in eq.(5), and

\begin{equation}
Q_{\ast }^{2}=\frac{GQ^{2}}{4\pi \epsilon _{0}c^{4}}.
\end{equation}

If $[x^{\mu }]=(t,r,\theta ,\phi )$, then the worldline $x^{\mu }(\tau )$,
where $\tau $ is the proper time along the path, of a particle moving in the
equatorial plane $\theta =\pi /2$ satisfies the equations [1,2]

\begin{equation}
\left( 1-\frac{\alpha }{r}+\frac{Q_{\ast }^{2}}{r^{2}}\right) \overset{\cdot
}{t}=\kappa ,
\end{equation}

\begin{equation}
c^{2}\left( 1-\frac{\alpha }{r}+\frac{Q_{\ast }^{2}}{r^{2}}\right) \overset{%
\cdot }{t}^{2}-\left( 1-\frac{\alpha }{r}+\frac{Q_{\ast }^{2}}{r^{2}}\right)
^{-1}\overset{\cdot }{r}^{2}-r^{2}\overset{\cdot }{\phi }^{2}=c^{2},
\end{equation}

\begin{equation}
r^{2}\overset{\cdot}{\phi}=h,
\end{equation}

where the derivative $\overset{\cdot }{}$ represents $d/d\tau $. The
coordinates $r$ and $\phi $ describe the position of the particle relative
to the charged star or black hole centered at the origin. The constant $h$
is identified as the angular momentum per unit rest mass of the particle,
and the constant $\kappa $ is identified to be the total energy per unit
rest energy of the particle

\begin{equation}
\kappa=\frac{E}{m_{0}c^{2}},
\end{equation}

where $E$ is the total energy of the particle in its orbit and $m_{0}$ is
the rest mass of the particle at $r=\infty $. Substituting eqs.(9) and (11)
into (10) gives the 'combined' energy equation

\begin{equation}
\overset{\cdot }{r}^{2}+c^{2}+\frac{h^{2}}{r^{2}}\left( 1-\frac{\alpha }{r}+%
\frac{Q_{\ast }^{2}}{r^{2}}\right) -\frac{c^{2}\alpha }{r}=c^{2}\kappa ^{2}.
\end{equation}

Substituting $dr/d\tau =(dr/d\phi )(d\phi /d\tau )=(h/r^{2})(dr/d\phi )$
into the combined energy equation gives the differential equation for the
trajectory of the particle

\begin{equation}
\left( \frac{du}{d\phi }\right) ^{2}=-Q_{\ast }^{2}u^{4}+\alpha
u^{3}-u^{2}\left( 1+\frac{c^{2}}{h^{2}}Q_{\ast }^{2}\right) +\left( \frac{2GM%
}{h^{2}}\right) u+\frac{c^{2}(\kappa ^{2}-1)}{h^{2}},
\end{equation}

where $u=1/r$. We define the dimensionless inverse distance $U$ from the
following relation

\begin{equation}
\frac{1}{q}\equiv \frac{\alpha }{r}=\alpha u\equiv \frac{1}{3}+4U.
\end{equation}

In terms of $U$, eq.(14) becomes

\begin{equation}
\left( \frac{dU}{d\phi }\right)
^{2}=a_{0}U^{4}+4a_{1}U^{3}+6a_{2}U^{2}+4a_{3}U+a_{4},
\end{equation}

where

\begin{equation}
a_{0}=-2^{4}\beta ^{2},
\end{equation}

\begin{equation}
a_{1}=1-\frac{4}{3}\beta ^{2},
\end{equation}

\begin{equation}
a_{2}=-\frac{1}{9}\beta ^{2}(1+6s^{2}),
\end{equation}

\begin{equation}
a_{3}=-\frac{1}{4}\left[ g_{2}+\frac{1}{3^{3}}\beta ^{2}(1+18s^{2})\right] ,
\end{equation}

\begin{equation}
a_{4}=-\left[ g_{3}+\frac{1}{2^{4}\cdot 3^{4}}\beta ^{2}(1+36s^{2})\right] ,
\end{equation}

where $\beta ^{2}$ is given by eq.(6) and where

\begin{align}
g_{2}& =\frac{1}{12}-s^{2}  \notag \\
g_{3}& =\frac{1}{216}+\frac{1}{6}s^{2}-\frac{1}{4}\kappa ^{2}s^{2}\equiv
\frac{1}{216}-\frac{1}{12}s^{2}+\frac{1}{4}(1-e^{2})s^{4},
\end{align}

$s^{2}$, $\kappa ^{2}$ and $e^{2}$ having been defined earlier in eqs.(2),
(12) and (1). The dimensionless inverse radial distance $U$ in place of $%
\alpha /r$ has been chosen so that, in the Schwarzschild limit $\beta ^{2}=0$%
, eq.(16) reduces to the corresponding equation studied in refs.3 and 4
given by

\begin{equation}
\left( \frac{dU}{d\phi }\right) ^{2}=4U^{3}-g_{2}U-g_{3}.
\end{equation}

Before we divide the parameter space $(e^{2},s^{2})$ for a given $\beta ^{2}$
into regions and present various analytic solutions of eq.(16) for the
trajectories of the particle, we note from the factor

\begin{equation*}
D\equiv 1-\frac{\alpha }{r}+\frac{Q_{\ast }^{2}}{r^{2}}=1-\frac{1}{q}+\frac{%
\beta ^{2}}{q^{2}}
\end{equation*}

that appears in eqs.(7), (9), (10), and (13) that positive $D$ means that
the coordinates $t$ and $r$ are timelike and spacelike respectively, whereas
negative $D$ means that the physical natures of the coordinates $t$ and $r$
are reversed. There are three different cases. (i) The case $\beta ^{2}>1/4$
leads to $D>0$ for all values of $q$. (ii) For $\beta ^{2}<1/4$, there are
two coordinate singularities occurring on the surfaces $q=q_{\pm }$ defined
by

\begin{equation*}
q_{\pm }=\frac{1}{2}\pm \left( \frac{1}{4}-\beta ^{2}\right) ^{1/2}.
\end{equation*}

The function $D$ is positive for $q>q_{+}$ or $q<q_{-}$, and is negative in
the region $q_{-}<q<q_{+}$. The case for $q=q_{+}$ can be compared to that
for the Schwarzschild horizon at $q=1$. (iii) For the case $\beta ^{2}=1/4$,
the function $D$ is positive everywhere except at $q=1/2$ where it equals
zero. The coordinate $q$ is spacelike everywhere except at $q=1/2$ and $%
q=1/2 $ is an event horizon.

We now discuss how the parameter space $(e^{2},s^{2})$ for a given $\beta
^{2}$ should be divided into regions for different types of solutions of
eq.(16). The discriminant $\Delta $ of the quartic equation

\begin{equation}
a_{0}U^{4}+4a_{1}U^{3}+6a_{2}U^{2}+4a_{3}U+a_{4}=0,
\end{equation}

where the $a^{\prime }s$ are given by eqs.(17)-(21), is given by

\begin{equation}
\Delta \equiv 27J^{2}-I^{3}
\end{equation}

where

\begin{equation}
I=a_{0}a_{4}-4a_{1}a_{3}+3a_{2}^{2},
\end{equation}

and

\begin{equation}
J=\left\vert
\begin{array}{ccc}
a_{0} & a_{1} & a_{2} \\
a_{1} & a_{2} & a_{3} \\
a_{2} & a_{3} & a_{4}%
\end{array}%
\right\vert ,
\end{equation}

For the quartic equation (24), $\Delta <0$ gives four real or four complex
roots, $\Delta >0$ gives two real and two complex roots, and $\Delta =0$
gives repeated roots. In the Schwarzschild limit $\beta ^{2}=0$, $\Delta $
reduces to $27g_{3}^{2}-g_{2}^{3}$. Analytic expressions for the roots of a
quartic equation can be written down but they are generally cumbersome and
we shall not give them here. The four roots can be numerically obtained for
a given set of parameters $\beta ^{2},e^{2},$ and $s^{2}$, but it would be
useful to know just from the given set of parameters represented by a point
in the parameter space what type of trajectories one would expect.

The mathematical conditions $\Delta <0$ and $\Delta >0$ divide the parameter
space $(e^{2},s^{2})$ for a given value of $\beta ^{2}$ into two regions
which we call Regions I and II respectively, with $\Delta =0$ forming the
boundary of the two regions. The three analytic solutions representing three
types of trajectories which we present below are all expressed in terms of
the Jacobian elliptic functions [5] with modulus $k$ that are analogous to
those we presented in refs.3 and 4 for the Schwarzschild geometry. The first
two solutions which we call Solutions (A1) and (A2) apply in Region I and
the third which we call Solution (B) applies in Region II. We do not discuss
the special motion along a radius for which $h=0$ and $s$ is infinite.

The boundary between Regions I and II given by the mathematical condition $%
\Delta =0$ is represented by three curves in the $(e^{2},s^{2})$ parameter
space for a given specific value of $\beta ^{2}$ which we shall refer to as
the $Y_{0}$, $Y_{1}^{\prime }$ and $Y_{1}$ curves respectively. Figure 1
shows the parameter space $(e^{2},s^{2})$ that exhibits part of the
horizontal $e^{2}$ axis from $-\infty $ to $+\infty $ and part of the
vertical $s^{2}$ axis from $0$ to $+\infty $. The $Y_{0}$ curve coincides
with the $e^{2}$-axis represented by $s^{2}=0$, is the bottom line of Region
I for $0\leq e^{2}\leq +\infty $, and is the bottom line of Region II for $%
-\infty \leq e^{2}<0$, for all values of $\beta ^{2}$. That is

\begin{equation*}
Y_{0}:s^{2}=0.
\end{equation*}

In Fig.1, the light dashed and light solid curves from $V_{0}$ are the $%
Y_{1}^{\prime }$ and $Y_{1}$ curves for the specific value of $\beta ^{2}=0$%
, and the heavy dashed and heavy solid curves from $V_{1}$ are the $%
Y_{1}^{\prime }$ and $Y_{1}$ curves for the specific value of $\beta
^{2}=1/4 $. Region I for a specific value of $\beta ^{2}$ is the parameter
space enclosed by $Y_{1}^{\prime }$ on the left, $Y_{1}$ on the top, and the
$e^{2} $-axis from $0$ to $+\infty $ at the bottom. Region II is the
parameter space outside of Region I above the $e^{2}$-axis. We describe all
this in greater detail below.

For a given value of $\beta ^{2}$, the $Y_{1}^{\prime }$ curve is
characterized by $\Delta =0$ and $k^{2}=0$, where $k$ is the modulus of the
Jacobian elliptic functions used to describe the trajectories of the
particle [see eqs.(34) and (46) below]. It extends from a special point that
we call the vertex point $(e_{v}^{2}(\beta ^{2}),s_{v}^{2}(\beta ^{2}))$ to
the origin $(0,0)$ and forms the left boundary of Region I with Region II.
The $Y_{1}$ curve is characterized by $\Delta =0$ and $k^{2}=1$. It extends
from $(e_{v}^{2}(\beta ^{2}),s_{v}^{2}(\beta ^{2}))$ to $(+\infty ,0)$ and
forms the upper boundary of Region I with Region II. The vertex point is in
fact the intersection point of all $k^{2}=const.$ curves for $0\leq
k^{2}\leq 1$ [see refs.3 and 4]. We write the $Y_{1}^{\prime }$ and $Y_{1}$
curves generally as

\begin{eqnarray*}
Y_{1}^{\prime } &:&s^{2}=y_{1}^{\prime }(e^{2},\beta ^{2}), \\
Y_{1} &:&s^{2}=y_{1}(e^{2},\beta ^{2}),
\end{eqnarray*}

where $y_{1}^{\prime }$ and $y_{1}$ are two specific functions of $e^{2}$
and $\beta ^{2}$ obtained from setting the discriminant $\Delta $ given by
eq.(25) of the quartic equation (24) equal to zero.

For the special value of $\beta ^{2}=0$, the Schwarzschild limit, we have
the simple expressions for $y_{1}^{\prime }$ and $y_{1}$ [3,4]

\begin{equation}
s^{2}=\frac{1-9e^{2}\mp \sqrt{(1+3e^{2})^{3}}}{27(1-e^{2})^{2}}
\end{equation}

with the upper and lower signs for $Y_{1}^{\prime }$ and $Y_{1}$
respectively. They meet and terminate at the vertex point $%
V_{0}=(e_{v}^{2},s_{v}^{2})=($ $-1/3,1/12)$ (see Fig.1). Equation (28) can
be inverted to give $e^{2}$ in terms of $s^{2}$ as

\begin{equation}
e^{2}=\frac{1-18s^{2}+54s^{4}\mp \sqrt{(1-12s^{2})^{3}}}{54s^{4}},
\end{equation}

For the special value of $\beta ^{2}=1/4$, we have the following simple
expressions for $y_{1}^{\prime }$ and $y_{1}$:

\begin{equation}
s^{2}=\frac{1-12e^{2}\mp \sqrt{(1+4e^{2})^{3}}}{2(3-4e^{2})^{2}}
\end{equation}

with the upper and lower signs for $Y_{1}^{\prime }$ and $Y_{1}$
respectively. They meet and terminate at the vertex point $%
V_{1}=(e_{v}^{2},s_{v}^{2})=($ $-1/4,1/8)$ (see Fig.1). Equation (30) can be
inverted to give $e^{2}$ in terms of $s^{2}$ as

\begin{equation}
e^{2}=\frac{1-12s^{2}+24s^{4}\mp \sqrt{(1-8s^{2})^{3}}}{32s^{4}}.
\end{equation}

We shall refer to $Y_{1}^{\prime }$ and $Y_{1}$ as the left and upper
boundaries respectively of Region I with Region II. The intersection point
of $Y_{1}^{\prime }$ and $Y_{1}$ is the vertex point where the innermost
stable circular orbit (ISCO) occurs (see the following section).

The left and upper boundary curves of Region I with Region II for the
intermediate values of $0<\beta ^{2}<1/4$ are between these two cases and
can be put in place approximately. The points on the curves $Y_{1}^{\prime }$
and $Y_{1}$ will be referred to as $(e_{1}^{\prime 2},s_{1}^{\prime 2})$ and
$(e_{1}^{2},s_{1}^{2})$ respectively.

The vertex points for various values of $\beta ^{2}$ can be conveniently
obtained numerically from setting $I=J=0$ from eqs.(26) and (27), and some
of these are given in the following table.

\begin{equation*}
\begin{array}{ccc}
\beta ^{2} & e_{v}^{2}(\beta ^{2}) & s_{v}^{2}(\beta ^{2}) \\
0.25 & -0.250000 & 0.125000 \\
0.20 & -0.267596 & 0.111453 \\
0.15 & -0.285240 & 0.101821 \\
0.10 & -0.302119 & 0.094384 \\
0.05 & -0.318135 & 0.088363 \\
0 & -0.333333 & 0.083333%
\end{array}%
\end{equation*}

Another "boundary" curve of interest is one that represents the total energy
equal to zero, or $\kappa ^{2}=0$ which in our $(e^{2},s^{2})$ parameter
space is represented by the curve $s^{2}(1-e^{2})=1$, or

\begin{equation}
Y_{2}:s^{2}=\frac{1}{1-e^{2}}.
\end{equation}

We shall refer to this curve as $Y_{2}$ [not shown in Fig.1] and refer to
eq.(32) as $s^{2}=y_{2}(e^{2})$ that applies to any value of $\beta ^{2}$ as
the top boundary of Region II. The points on this boundary will be referred
to as $(e_{2}^{2},s_{2}^{2})$. The physical requirement that $\kappa
^{2}\geq 0$, where $\kappa $ is given by eq.(12), leads to the condition
that $s^{2}\leq 1/(1-e^{2})$. We call the region $s^{2}>s_{2}^{2}$ Region
II'.

The analytic solutions that we shall present for the R-N geometry, in
analogy with those we presented in refs.3 and 4 for the Schwarzschild
geometry, are given in three forms which we call Solution (A1), Solution
(A2) and Solution (B); the first two apply in Region I and the third applies
in Region II. When $\Delta \leq 0$, the four real roots of the quartic
equation (24) are arranged in the order $\varepsilon _{1}>\varepsilon
_{2}>\varepsilon _{3}>\varepsilon _{4}$. All analytic solutions of eq.(16)
are expressed in terms of the Jacobian elliptic functions [5] which become
circular or hyperbolic functions for special cases.

Solution (A1) For $\Delta \leq 0$, $\varepsilon _{1}>\varepsilon
_{2}>\varepsilon _{3}\geq U>\varepsilon _{4}$ Applicable in Region I

Writing the right-hand side of eq.(16) as $16\beta ^{2}(\varepsilon
_{1}-U)(\varepsilon _{2}-U)(\varepsilon _{3}-U)(U-\varepsilon _{4})$, we
find the equation for the trajectory to be

\begin{equation}
\frac{1}{q}=\frac{1}{3}+4\frac{\varepsilon _{4}(\varepsilon _{1}-\varepsilon
_{3})+\varepsilon _{1}(\varepsilon _{3}-\varepsilon _{4})sn^{2}(\gamma \phi
,k)}{(\varepsilon _{1}-\varepsilon _{3})+(\varepsilon _{3}-\varepsilon
_{4})sn^{2}(\gamma \phi ,k)},
\end{equation}

where the modulus $k$ of the elliptic functions is given by

\begin{equation}
k^{2}=\frac{(\varepsilon _{1}-\varepsilon _{2})(\varepsilon _{3}-\varepsilon
_{4})}{(\varepsilon _{1}-\varepsilon _{3})(\varepsilon _{2}-\varepsilon _{4})%
},
\end{equation}

and

\begin{equation}
\gamma =2\beta \sqrt{(\varepsilon _{1}-\varepsilon _{3})(\varepsilon
_{2}-\varepsilon _{4})}.
\end{equation}

The modulus $k$ has a range $0\leq k^{2}\leq 1$. For $k^{2}=0$, $%
sn(x,0)=\sin x$, and for $k^{2}=1$, $sn(x,1)=\tanh x$.

For the bound orbits of the elliptic-type in Region I confined to $e^{2}<1$,
$\varepsilon _{4}$ is greater than $-1/12$, and the initial point at $\phi
=0 $ of the bound orbit has been chosen so that the dimensionless distance $%
q $ of the particle from the black hole is given by $1/q=1/q_{\max
}=1/3+4\varepsilon _{4}$, where $q_{\max }$ is the maximum distance of the
particle from the black hole. The minimum distance $q_{\min }$ of the
particle from the black hole is given by $1/q_{\min }=1/3+4\varepsilon _{3}$%
. The geometric eccentricity can be defined in terms of $q_{\max }$ and $%
q_{\min }$ by $(q_{\max }-q_{\min })/(q_{\max }+q_{\min })$. The
precessional angle $\Delta \phi $ is given by

\begin{equation}
\Delta \phi =\frac{2K(k)}{\gamma }-2\pi ,
\end{equation}

where $K(k)$ is the complete elliptic integral of the first kind. For the
unbound parabolic-type orbits characterized by $e^{2}=1$ in Region I, $%
\varepsilon _{4}=-1/12$, and for the unbound hyperbolic-type orbits
characterized by $e^{2}>1$ in Region I, $\varepsilon _{4}<-1/12$, the
particle comes from infinity from a polar angle $\Psi _{1}$ given by

\begin{equation}
\Psi _{1}=\gamma ^{-1}sn^{-1}\left( \sqrt{-\frac{(\varepsilon
_{1}-\varepsilon _{3})(1+12\varepsilon _{4})}{(\varepsilon _{3}-\varepsilon
_{4})(1+12\varepsilon _{1})}},k\right) ,
\end{equation}

and returns to infinity at a polar angle $\Psi _{1}^{\prime }$ given by

\begin{equation}
\Psi _{1}^{\prime }=\frac{2K}{\gamma }-\Psi _{1},
\end{equation}

Equation (37) is obtained from eq.(33) by setting $q=\infty $. The
deflection angle $\Theta $ can be defined as

\begin{equation}
\Theta \equiv \Psi _{1}^{\prime }-\Psi _{1}-\pi .
\end{equation}

Solution (A2) For $\Delta \leq 0$, $\varepsilon _{1}>U>\varepsilon
_{2}>\varepsilon _{3}>\varepsilon _{4}$ Applicable in Region I

Writing the right-hand side of eq.(16) as $16\beta ^{2}(\varepsilon
_{1}-U)(U-\varepsilon _{2})(U-\varepsilon _{3})(U-\varepsilon _{4})$, we
find the equation for the trajectory to be

\begin{equation}
\frac{1}{q}=\frac{1}{3}+4\frac{\varepsilon _{2}(\varepsilon _{1}-\varepsilon
_{3})-\varepsilon _{3}(\varepsilon _{1}-\varepsilon _{2})sn^{2}(\gamma \phi
,k)}{(\varepsilon _{1}-\varepsilon _{3})-(\varepsilon _{1}-\varepsilon
_{2})sn^{2}(\gamma \phi ,k)},
\end{equation}

where $k$ and $\gamma $ are given by the same eqs.(34) and (35). The initial
distance $q_{0}$ of the particle from the black hole at $\phi =0$ has been
chosen to be given by $1/q_{0}=1/3+4\varepsilon _{2}$. It is clear that $%
q_{0}$ is less than $q_{\min }$ given by Solution (A1) at any given point
inside Region I.

Solution (B) For $\Delta >0$, $\varepsilon _{1}\geq U>\varepsilon _{2}$, and
$\varepsilon _{3},\bar{\varepsilon}_{3}$ complex, where $\varepsilon
_{1},\varepsilon _{2},\varepsilon _{3}$ and $\bar{\varepsilon}_{3}$ are the
roots of the quartic equation (24).

This Solution is applicable in Region II.

Writing the right-hand side of eq.(16) as $16\beta ^{2}(\varepsilon
_{1}-U)(U-\varepsilon _{2})(U-\varepsilon _{3})(U-\bar{\varepsilon}_{3})$,
we find the equation for the trajectory to be

\begin{equation}
\frac{1}{q}=\frac{1}{3}+4\frac{\varepsilon _{2}A(1+cn(2\gamma \phi
,k))+\varepsilon _{1}B(1-cn(2\gamma \phi ,k))}{A(1+cn(2\gamma \phi
,k))+B(1-cn(2\gamma \phi ,k))},
\end{equation}

where

\begin{equation}
A^{2}=(\varepsilon _{1}-b_{1})^{2}+a_{1}^{2},
\end{equation}

\begin{equation}
B^{2}=(\varepsilon _{2}-b_{1})^{2}+a_{1}^{2},
\end{equation}

\begin{equation}
a_{1}^{2}=-\frac{(\varepsilon _{3}-\bar{\varepsilon}_{3})^{2}}{4},
\end{equation}

\begin{equation}
b_{1}=\frac{\varepsilon _{3}+\bar{\varepsilon}_{3}}{2},
\end{equation}

\begin{equation}
k^{2}=\frac{(\varepsilon _{1}-\varepsilon _{2})^{2}-(A-B)^{2}}{4AB},
\end{equation}

and

\begin{equation}
\gamma =2\beta \sqrt{AB}.
\end{equation}

For $k^{2}=0$, $cn(x,0)=\cos x$, and for $k^{2}=1$, $cn(x,1)=\sec hx$. For $%
e^{2}<1$ in Region II, $\varepsilon _{2}$ is greater than $-1/12$. The
initial distance $q$ at $\phi =0$ of the particle from the black hole has
been chosen to be given by $1/q_{0}=1/3+\varepsilon _{2}$. For $e^{2}\geq 1$
in Region II, $\varepsilon _{2}$ is $\leq -1/12$, the particle comes from
infinity at a polar angle $\Phi _{2}$ given by

\begin{equation}
\Phi _{2}=(2\gamma )^{-1}cn^{-1}\left( -\frac{(1+12\varepsilon
_{2})A+(1+12\varepsilon _{1})B}{(1+12\varepsilon _{2})A-(1+12\varepsilon
_{1})B},k\right) ,
\end{equation}

and returns to infinity at a polar angle $\Phi _{2}^{\prime }$ given by

\begin{equation}
\Phi _{2}^{\prime }=\frac{2K}{\gamma }-\Phi _{2}.
\end{equation}

In all three Solutions above, the Schwarzschild limit is obtained if we let $%
\beta \rightarrow 0,$ $\varepsilon _{1}\rightarrow \infty ,$ such that $%
2\beta \sqrt{\varepsilon _{1}}\rightarrow 1$.

\section{Examples of the Particle Trajectories Including Special Cases}

Region I $(\Delta \leq 0)$ allows two types of trajectory, given by
Solutions (A1) and (A2) respectively, depending on the initial distance of
the particle from the black hole. Solution (A1) expressed by eq.(33) gives
trajectories that are similar to those for the Schwarzschild geometry which
we classify into three types: the elliptic type $(e_{1}^{\prime 2}\leq
e^{2}<1)$, the parabolic type $(e^{2}=1)$ and the hyperbolic type $(e^{2}>1)$
[4, 20].

Figures 2-5 show some examples of trajectories of an electrically neutral
particle when $q$ is plotted versus $\phi $. These trajectories are given by
eqs.(33), (40) and (41) where the electrically charged black hole is located
at the origin.

Figure 2 shows an example of a precessing elliptic-type orbit in Region I
given by eq.(33) for $e^{2}=1/4$, $s^{2}=0.037725$, and $\beta ^{2}=1/4$ for
which the precession angle $\Delta \phi $ given by eq.(36) is $43.0^{\circ }$%
. Comparing this with the elliptic-type orbit at the same point $%
(e^{2},s^{2})$ for the case $\beta ^{2}=0$ (the Schwarzschild limit) for
which $\Delta \phi =60.5^{\circ }$ [3, Fig.5], we see that the presence of
electric charge on the black hole reduces the precession angle $\Delta \phi $%
.

Figure 3 shows an example of a hyperbolic-type orbit in Region I given by
eq.(33) for $e^{2}=4$, $s^{2}=0.029564$, and $\beta ^{2}=1/4$ for which the
deflection angle $\Theta $ given by eq.(39) is $101.1^{\circ }$. Comparing
this with the hyperbolic-type orbit at the same point $(e^{2},s^{2})$ for
the case $\beta ^{2}=0$ for which $\Theta =115.6^{\circ }$ [20, Fig.3], we
see that the presence of electric charge on the black hole reduces the
deflection angle $\Theta $.

The presence of electric charge on the black hole makes a greater difference
to the trajectories given by eq.(40) of Solution (A2) in Region I and to
those given by eq.(41) of Solution (B) in Region II. It changes a
terminating orbit into a "wandering" orbit in which the particle wanders
around the black hole as its distance $q$ from the black hole oscillates
between $1/\varepsilon _{2}$ and $1/\varepsilon _{1}$. Figure 4 shows an
example of a wandering orbit in Region I given by eq.(40) for $e^{2}=1/4$, $%
s^{2}=0.0319276$ and $\beta ^{2}=0.1$ for which $q_{+}=0.88730$ and $%
q_{-}=0.11270$. The trajectory, starting at a distance $q_{0}=1.0178$ enters
a region between $q_{+}$ and $q_{-}$ and does not terminate at the center of
the black hole but instead it wanders around the center of the black hole.
The physical significance of the trajectory in the region $q_{-}<q<q_{+}$ is
not completely clear [1,2]. Compare this with the terminating orbit for the
case $\beta ^{2}=0$ [3, Fig.6(b)].

Figure 5 shows an example of a wandering orbit given by eq.(41) in Region II
for $e^{2}=4$, $s^{2}=0.040$ and $\beta ^{2}=1/4$. The particle comes from
infinity at an angle of $55.3^{\circ }$ and instead of terminating at the
black hole as for the case of $\beta ^{2}=0$, it swirls around the black
hole and goes to infinity at an angle of $754.6^{\circ }$. It, however, goes
through a region of $q<q_{+}=q_{-}=0.5$ to a minimum distance of $q_{\min
}=0.3965$ the physical significance of which is not clear, and thus the part
of the trajectory going back to infinity may not be physically realized.

The wandering orbits given by eqs.(40) and (41) become terminating orbits in
the Schwarzschild limit $\beta \rightarrow 0$, $\varepsilon _{1}\rightarrow
\infty $ such that $2\beta \sqrt{\varepsilon _{1}}\rightarrow 1$.

The boundaries $Y_{0}$, $Y_{1}^{\prime }$ and $Y_{1}$ represent the special
case $\Delta =0$ on which the particle trajectories are of special types
[see Fig.1 and the description given by eqs.(28)-(31)].

On $Y_{0}$, we have $s^{2}=0$. The quartic equation (24) becomes

\begin{equation*}
-\left( U+\frac{1}{12}\right) ^{2}\left[ 16\beta ^{2}U^{2}-\frac{4}{3}\left(
-2\beta ^{2}+3\right) U+\frac{1}{9}\left( \beta ^{2}+6\right) \right] =0
\end{equation*}

which, for $0\leq \beta ^{2}\leq 1/4$, has four real roots given by

\begin{eqnarray*}
\varepsilon _{1} &=&\frac{1}{24}\frac{-2\beta ^{2}+3+3\sqrt{1-4\beta ^{2}}}{%
\beta ^{2}}, \\
\varepsilon _{2} &=&\frac{1}{24}\frac{-2\beta ^{2}+3-3\sqrt{1-4\beta ^{2}}}{%
\beta ^{2}}, \\
\varepsilon _{3} &=&\varepsilon _{4}=-\frac{1}{12}.
\end{eqnarray*}

The trajectory of the particle is given from eq.(33) by $1/q=0$ or $q=\infty
$, i.e. the particle is infinitely far away from the black hole and its
trajectory is a straight line independent of $\beta ^{2}$. From eq.(34), the
curve $Y_{0}$ is also characterized by $k^{2}=0$.

On the left boundary $Y_{1}^{\prime }$ (or $s_{1}^{\prime 2}$) of Region I,
we have $\varepsilon _{3}=\varepsilon _{4}$ and $k^{2}=0$, the elliptic-type
orbits given by Solution (A1) become stable circular orbits with radii $%
q_{c} $ given by

\begin{equation}
\frac{1}{q_{c}}=\frac{1}{3}+4\varepsilon _{4},
\end{equation}

whereas the wandering orbits given by Solution (A2) continue to be wandering
orbits given by

\begin{equation}
\frac{1}{q}=\frac{1}{3}+4\frac{\varepsilon _{2}(\varepsilon _{1}-\varepsilon
_{3})-\varepsilon _{3}(\varepsilon _{1}-\varepsilon _{2})\sin ^{2}(\gamma
\phi )}{(\varepsilon _{1}-\varepsilon _{3})-(\varepsilon _{1}-\varepsilon
_{2})\sin ^{2}(\gamma \phi )}.
\end{equation}

On the upper boundary $Y_{1}$ (or $s_{1}^{2}$) of Region I where $%
\varepsilon _{2}=\varepsilon _{3}$ and $k^{2}=1$, the elliptic-, parabolic-
and hyperbolic-type orbits given by Solution (A1) and the wandering orbits
given by Solution (A2) all become unstable asymptotic circular orbits with
radii $q_{u}$ given by

\begin{equation}
\frac{1}{q_{u}}=\frac{1}{3}+4\varepsilon _{3}.
\end{equation}

The intersection of $Y_{1}^{\prime }$ and $Y_{1}$ in the parameter space is
the vertex point $V$ where $\varepsilon _{1}=\varepsilon _{2}=\varepsilon
_{3}$ and where the innermost stable circular orbit (ISCO) occurs. For the
Schwarzschild geometry ($\beta ^{2}=0$), $V$ is at $%
(e^{2},s^{2})=(-1/3,1/12) $ or $(\kappa ^{2},s^{2})=(8/9,1/12)$ shown as $%
V_{0}$ in Fig.1, and the radius of ISCO is $q_{c}=3$ or $r_{c}=6GM/c^{2}$.
For the R-N geometry at the special value of $\beta ^{2}=1/4$, $V$ is at $%
(e^{2},s^{2})=(-1/4,1/8)$ or $(\kappa ^{2},s^{2})=(27/32,1/8)$ [21] shown as
$V_{1}$ in Fig.1, and the radius of ISCO is $q_{c}=2$ or $r_{c}=4GM/c^{2}$.

The upper boundary $Y_{2}$ of Region II is characterized by zero total
energy or $\kappa ^{2}=0$. The quartic equation (24) becomes

\begin{equation*}
-\left( U^{2}+\frac{1}{6}U+\frac{1}{12^{2}}+\frac{1}{4}s^{2}\right) \left[
16\beta ^{2}U^{2}+\frac{4}{3}\left( 2\beta ^{2}-3\right) U+\frac{1}{9}\left(
\beta ^{2}+6\right) \right] =0
\end{equation*}

which, for $0\leq \beta ^{2}\leq 1/4$, has four roots given by

\begin{eqnarray*}
\varepsilon _{1} &=&\frac{1}{24}\frac{-2\beta ^{2}+3+3\sqrt{1-4\beta ^{2}}}{%
\beta ^{2}}, \\
\varepsilon _{2} &=&\frac{1}{24}\frac{-2\beta ^{2}+3-3\sqrt{1-4\beta ^{2}}}{%
\beta ^{2}}, \\
\varepsilon _{3} &=&-\frac{1}{12}+\frac{1}{2}is, \\
\bar{\varepsilon}_{3} &=&-\frac{1}{12}-\frac{1}{2}is,
\end{eqnarray*}

where $\beta ^{2}$ is assumed to be $\leq $ $1/4$. The trajectory is given
by eq.(41) with $k^{2}$ and $\gamma $ given by

\begin{eqnarray*}
k^{2} &=&\frac{1}{2}-\frac{1}{2}\frac{1+4\beta ^{2}s^{2}}{\left[
1+4(1-2\beta ^{2})s^{2}+16\beta ^{4}s^{4}\right] ^{1/2}}, \\
\gamma &=&\frac{1}{2}\left[ 1+4(1-2\beta ^{2})s^{2}+16\beta ^{4}s^{4}\right]
^{1/4}.
\end{eqnarray*}

The initial distance $q_{0}$ of the particle from the black hole at $\phi =0$
is given by $1/q_{0}=1/3+4\varepsilon _{2}$ which yields

\begin{equation*}
q_{0}=\frac{1}{2}\left( 1+\sqrt{1-4\beta ^{2}}\right)
\end{equation*}

which is the distance $q_{+}$ of the outer horizon from the black hole. Thus
all particle trajectories on $Y_{2}$ start from the horizon (in a direction
perpendicular to the line joining it to the center of the black hole).

\section{Light Trajectories in the R-N Geometry}

The trajectory of a photon is a null geodesic. Instead of using the proper
time $\tau $ as a parameter, we use some affine parameter $\sigma $ along
the geodesic. Considering motion in the equatorial plane, the equations of
motion in the R-N geometry are given by

\begin{equation}
\left( 1-\frac{\alpha }{r}+\frac{Q_{\ast }^{2}}{r^{2}}\right) \overset{\cdot
}{t}=\kappa ,
\end{equation}

\begin{equation}
c^{2}\left( 1-\frac{\alpha }{r}+\frac{Q_{\ast }^{2}}{r^{2}}\right) \overset{%
\cdot }{t}^{2}-\left( 1-\frac{\alpha }{r}+\frac{Q_{\ast }^{2}}{r^{2}}\right)
^{-1}\overset{\cdot }{r}^{2}-r^{2}\overset{\cdot }{\phi }^{2}=0,
\end{equation}

\begin{equation}
r^{2}\overset{\cdot}{\phi}=h,
\end{equation}

where the derivative $\overset{\cdot }{}$ represents $d/d\sigma $. The
analog of eq.(13) is

\begin{equation}
\overset{\cdot }{r}^{2}+\frac{h^{2}}{r^{2}}\left( 1-\frac{\alpha }{r}+\frac{%
Q_{\ast }^{2}}{r^{2}}\right) =c^{2}\kappa ^{2}.
\end{equation}

Substituting $dr/d\sigma =(dr/d\phi )(d\phi /d\sigma )=(h/r^{2})(dr/d\phi )$
into the combined energy equation above gives the differential equation for
the trajectory of light

\begin{equation}
\left( \frac{du}{d\phi }\right) ^{2}=-Q_{\ast }^{2}u^{4}+\alpha u^{3}-u^{2}+%
\frac{c^{2}\kappa ^{2}}{h^{2}},
\end{equation}

where $u=1/r$. We define the dimensionless inverse distance $U$ by

\begin{equation}
U=\frac{1}{q}\equiv \frac{\alpha }{r}=\alpha u.
\end{equation}

In terms of $U$, eq.(57) becomes

\begin{equation}
\left( \frac{dU}{d\phi }\right) ^{2}=-\beta ^{2}U^{4}+U^{3}-U^{2}+\frac{%
c^{2}\kappa ^{2}\alpha ^{2}}{h^{2}},
\end{equation}

where $\beta ^{2}$ is defined by eq.(6) as before, but $U$ is defined
somewhat differently from eq.(15). We note that the trajectory depends on
the ratio $\kappa /h$ (and $\beta $) and not on $\kappa $ and $h$ separately
[19]. We could use the four roots of the quartic equation

\begin{equation}
-\beta ^{2}U^{4}+U^{3}-U^{2}+\frac{c^{2}\kappa ^{2}\alpha ^{2}}{h^{2}}=0
\end{equation}

obtained numerically from the given set of parameters $\beta ^{2}$ and $%
\kappa /h$ for characterizing the trajectory. Instead, as we did in a
similar fashion in ref.3, we replace $\kappa /h$ by an alternative parameter
as follows. Let $R$ denote the distance of the light beam to the center of
the black hole when the trajectory of the light beam is such that $dU/d\phi
=0$. When a light beam is simply deflected by the presence of the charged
black hole, $R$ is the closest distance of the light beam to the black hole
and may be unique; but for a more general trajectory it may not be unique.
We let $U_{1}=R/\alpha $ denote the value of $U$ at which $dU/d\phi =0$.
Then we can replace $c^{2}\kappa ^{2}\alpha ^{2}/h^{2}$ by $\beta
^{2}U_{1}^{4}-U_{1}^{3}+U_{1}^{2}$ and write eq.(59) as

\begin{equation}
\left( \frac{dU}{d\phi }\right) ^{2}=-\beta ^{2}U^{4}+U^{3}-U^{2}+\beta
^{2}U_{1}^{4}-U_{1}^{3}+U_{1}^{2}.
\end{equation}

We do not discuss the special case when the light is along a path that is
directly toward the black hole. The advantage of using eq.(61) is that one
root $U=U_{1}$ of the quartic equation (60) is assumed known or given
physically, and the other three roots of the resulting cubic equation

\begin{equation}
U^{3}+a_{1}U^{2}+a_{2}U+a_{3}=0
\end{equation}

where

\begin{equation}
a_{1}=-(\beta ^{-2}-U_{1}),
\end{equation}

\begin{equation}
a_{2}=(1-U_{1})\beta ^{-2}+U_{1}^{2},
\end{equation}

\begin{equation}
a_{3}=[(1-U_{1})\beta ^{-2}+U_{1}^{2}]U_{1},
\end{equation}

can be written down analytically in rather simple expressions. Defining

\begin{equation}
a=-\frac{\beta ^{-4}}{3}\left\{ 1-(3-U_{1})\beta ^{2}-2U_{1}^{2}\beta
^{4}\right\} ,
\end{equation}

\begin{equation}
b=-\frac{2\beta ^{-6}}{27}\left\{ 1-\frac{3(3-U_{1})}{2}\beta ^{2}-\frac{%
3(6U_{1}-5U_{1}^{2})}{2}\beta ^{4}-10U_{1}^{3}\beta ^{6}\right\} ,
\end{equation}

the discriminant $\Delta $ of the cubic equation (62) is given by

\begin{equation}
\Delta =\frac{b^{2}}{4}+\frac{a^{3}}{27},  \notag
\end{equation}

or

\begin{equation}
\Delta =-\frac{\beta ^{-8}}{2^{2}\cdot 3^{3}}\left\{
\begin{array}{c}
(1+2U_{1}-3U_{1}^{2})-2(2+4U_{1}-9U_{1}^{2}+U_{1}^{3})\beta ^{2} \\
-(20-12U_{1}+3U_{1}^{2})U_{1}^{2}\beta ^{4}-8(4-3U_{1})U_{1}^{4}\beta
^{6}-16U_{1}^{6}\beta ^{8}%
\end{array}%
\right\} .
\end{equation}

For $\Delta \leq 0$, the cubic equation (62) has three real roots. We define

\begin{equation}
\cos \theta =-\frac{b}{2\sqrt{-\frac{a^{3}}{27}}}.
\end{equation}

The three real roots of the cubic equation (62) are given by

\begin{equation}
x_{1}=2\sqrt{\frac{-a}{3}}\cos \frac{\theta }{3}+\frac{1}{3}\left( \beta
^{-2}-U_{1}\right) ,
\end{equation}

\begin{equation}
x_{2}=2\sqrt{\frac{-a}{3}}\cos \frac{\theta +4\pi }{3}+\frac{1}{3}\left(
\beta ^{-2}-U_{1}\right) ,
\end{equation}

\begin{equation}
x_{3}=2\sqrt{\frac{-a}{3}}\cos \frac{\theta +2\pi }{3}+\frac{1}{3}\left(
\beta ^{-2}-U_{1}\right) .
\end{equation}

For $\Delta >0$, the cubic equation (62) has one real and two complex roots
given by

\begin{equation}
x_{1}=\left( -\frac{b}{2}+\sqrt{\Delta }\right) ^{1/3}+\left( -\frac{b}{2}-%
\sqrt{\Delta }\right) ^{1/3}+\frac{1}{3}\left( \beta ^{-2}-U_{1}\right) ,
\end{equation}

\begin{equation}
x_{2}=\omega \left( -\frac{b}{2}+\sqrt{\Delta }\right) ^{1/3}+\omega
^{2}\left( -\frac{b}{2}-\sqrt{\Delta }\right) ^{1/3}+\frac{1}{3}\left( \beta
^{-2}-U_{1}\right) ,
\end{equation}

\begin{equation}
x_{3}==\overline{x}_{2}=\omega ^{2}\left( -\frac{b}{2}+\sqrt{\Delta }\right)
^{1/3}+\omega \left( -\frac{b}{2}-\sqrt{\Delta }\right) ^{1/3}+\frac{1}{3}%
\left( \beta ^{-2}-U_{1}\right) ,
\end{equation}

where $\omega =-1/2+i\sqrt{3}/2$ and $\omega ^{2}=-1/2-i\sqrt{3}/2$.

For the Schwarzschild geometry, the single parameter $U_{1}$ can be used to
characterize the trajectory of light, and there are three regions: Region I
for $0\leq U_{1}\leq 2/3$ $(\infty >R\geq 3GM/c^{2})$, Region II for $%
2/3<U_{1}\leq 1$ $(3GM/c^{2}>R\geq 2GM/c^{2})$, and Region III for $%
1<U_{1}\leq \infty $ $(2GM/c^{2}>R\geq 0)$.

For the R-N geometry, the parameter space $(U_{1},\beta ^{2})$ can be
divided also into basically three regions as shown in Fig.6. The
mathematical condition $\Delta =0$ produces three curves which we call $%
y_{1},y_{2}$ and $y_{3}$.

\begin{equation}
y_{1}:\beta ^{2}=\frac{1}{12U_{1}^{2}}\left( f+\frac{19U_{1}^{2}-52U_{1}+4}{f%
\text{ }}+2U_{1}-4\right) ,
\end{equation}

where

\begin{equation}
f=\left\{
\begin{array}{c}
3\sqrt{6U_{1}}\left(
729U_{1}^{5}-486U_{1}^{4}-54U_{1}^{3}+1144U_{1}^{2}+736U_{1}+128\right)
^{1/2} \\
+215U_{1}^{3}-192U_{1}^{2}+276U_{1}+8%
\end{array}%
\right\} ^{1/3};
\end{equation}

\begin{equation}
y_{2}:\beta ^{2}=\frac{3U_{1}-2}{4U_{1}^{2}};
\end{equation}

\begin{equation}
y_{3}:\beta ^{2}=\frac{1}{U_{1}}-\frac{1}{U_{1}^{2}};
\end{equation}

and they are shown in Fig.6. The curves $y_{1}$ and $y_{2}$ touch at what we
call the vertex point $V$ the coordinates of which are given by $%
(4/3,9/32)=(1.33333,0.28125)$. The curves $y_{2}$ and $y_{3}$ intersect at $%
(2,1/4)$.

Region I is the parameter space bounded by the $\beta ^{2}$-axis on the
left, by $y_{1}$ on the top [between $(0,1/4)$ and $V$], and by $y_{2}$ on
the right [between $(2/3,0)$ and $V$]. Region II is the parameter space
bounded by $y_{2}$ on the left [between $(2/3,0)$ and $V$], by $y_{3}$ on
the right [between $(1,0)$ and $(+\infty ,0)$], and by $y_{1}$ at the top
[between $V$ and $(+\infty ,0)$]. Region II is divided into two parts IIA
and IIB by the curve $y_{2}$ between $V$ and $(2,1/4)$. Region III is the
parameter space under the curve $y_{3}$, and it is divided into two parts
IIIA and IIIB by the curve $y_{2}$ from $(2,1/4)$ to $(+\infty ,0)$.

The analytic solutions that we shall present for the trajectory of light in
the R-N geometry, in analogy with those we presented in ref.3 for the
Schwarzschild geometry, are given in three forms for Regions I, II and III
respectively. The four roots of the quartic equation (60) consist of one
real root $U_{1}$ and three roots of the cubic equation (62). When $\Delta $
given by eq.(68) is $\leq 0$, the four real roots are arranged in the order $%
\varepsilon _{1}>\varepsilon _{2}>\varepsilon _{3}>\varepsilon _{4}$, and we
shall identify the four $\varepsilon ^{\prime }s$ with $U_{1}$ and the $%
x^{\prime }s$ given by eqs.(70)-(72) later. When $\Delta $ given by eq.(68)
is $>0$, there are two real roots $U_{1}$ and $x_{1}$, and two complex roots
$x_{2}$ and $\bar{x}_{2}$ given by eqs.(73)-(75).

Solution (I) For Region I $\Delta \leq 0$, $\varepsilon _{1}>\varepsilon
_{2}>\varepsilon _{3}\geq U>\varepsilon _{4}$. Here we set $\varepsilon
_{1}=x_{1},$ $\varepsilon _{2}=x_{2},$ $\varepsilon _{3}=U_{1},$ $%
\varepsilon _{4}=x_{3}$ from eqs.(70)-(72).

Writing the right-hand side of eq.(61) as $\beta ^{2}(\varepsilon
_{1}-U)(\varepsilon _{2}-U)(\varepsilon _{3}-U)(U-\varepsilon _{4})$, we
find the equation for the trajectory to be

\begin{equation}
\frac{1}{q}=\frac{\varepsilon _{3}(\varepsilon _{2}-\varepsilon
_{4})-\varepsilon _{2}(\varepsilon _{3}-\varepsilon _{4})sn^{2}(\gamma \phi
,k)}{(\varepsilon _{2}-\varepsilon _{4})-(\varepsilon _{3}-\varepsilon
_{4})sn^{2}(\gamma \phi ,k)},
\end{equation}

where the modulus $k$ of the elliptic functions is given by

\begin{equation}
k^{2}=\frac{(\varepsilon _{1}-\varepsilon _{2})(\varepsilon _{3}-\varepsilon
_{4})}{(\varepsilon _{1}-\varepsilon _{3})(\varepsilon _{2}-\varepsilon _{4})%
},
\end{equation}

and

\begin{equation}
\gamma =\frac{1}{2}\beta \sqrt{(\varepsilon _{1}-\varepsilon
_{3})(\varepsilon _{2}-\varepsilon _{4})}.
\end{equation}

Setting $U=0$ and $\phi =\pi /2+\Delta \phi /2$, where $\Delta \phi $ is the
total angle of deflection of the light beam [3], we find

\begin{equation}
sn^{2}\left[ \gamma \left( \frac{\pi }{2}+\frac{\Delta \phi }{2}\right) ,k%
\right] =\frac{(\varepsilon _{2}-\varepsilon _{4})\varepsilon _{3}}{%
(\varepsilon _{3}-\varepsilon _{4})\varepsilon _{2}},
\end{equation}

or

\begin{equation}
\Delta \phi =-\pi +\frac{2}{\gamma }sn^{-1}(\psi ,k),
\end{equation}

where

\begin{equation}
\psi =\sqrt{\frac{(\varepsilon _{2}-\varepsilon _{4})\varepsilon _{3}}{%
(\varepsilon _{3}-\varepsilon _{4})\varepsilon _{2}}.}
\end{equation}

Solution (II) For Region II $\Delta \leq 0$, $\varepsilon _{1}\geq
U>\varepsilon _{2}>\varepsilon _{3}>\varepsilon _{4}$. Write the right-hand
side of eq.(61) as $\beta ^{2}(\varepsilon _{1}-U)(U-\varepsilon
_{2})(U-\varepsilon _{3})(U-\varepsilon _{4})$.

For Region IIA bounded by $y_{2}$ [between $(2/3,0)$ and $(2,1/4)$] and $%
y_{3}$ [between $(1,0)$ and $(2,1/4)$], we set $\varepsilon _{1}=x_{1},$ $%
\varepsilon _{2}=U_{1},$ $\varepsilon _{3}=x_{2},$ $\varepsilon _{4}=x_{3}$
from eqs.(70)-(72). We find the equation for the trajectory to be

\begin{equation}
\frac{1}{q}=\frac{\varepsilon _{2}(\varepsilon _{1}-\varepsilon
_{3})-\varepsilon _{3}(\varepsilon _{1}-\varepsilon _{2})sn^{2}(\gamma \phi
,k)}{(\varepsilon _{1}-\varepsilon _{3})-(\varepsilon _{1}-\varepsilon
_{2})sn^{2}(\gamma \phi ,k)},
\end{equation}

where $k$ and $\gamma $ are given by the same eqs.(81) and (82), and where
we have chosen the initial value of $q$ at $\phi =0$ to be given by $%
1/q=\varepsilon _{2}$.

For Region IIB bounded by $y_{2}$ [between $V$ and $(2,1/4)$], $y_{3}$
[between $(2,1/4)$ and $(+\infty ,0)$], and $y_{1}$ [between $V$ and $%
(+\infty ,0)$], we set $\varepsilon _{1}=U_{1},$ $\varepsilon _{2}=x_{1},$ $%
\varepsilon _{3}=x_{2},$ $\varepsilon _{4}=x_{3}$ from eqs.(70)-(72). We
find the same equation for the trajectory as that given by eq.(86), or if we
choose the initial value at $\phi =0$ for $q$ to be $1/q=\varepsilon _{1}$,
we have the following trajectory

\begin{equation}
\frac{1}{q}=\frac{\varepsilon _{1}(\varepsilon _{2}-\varepsilon
_{4})+\varepsilon _{4}(\varepsilon _{1}-\varepsilon _{2})sn^{2}(\gamma \phi
,k)}{(\varepsilon _{2}-\varepsilon _{4})+(\varepsilon _{1}-\varepsilon
_{2})sn^{2}(\gamma \phi ,k)}.
\end{equation}

The two expressions for the trajectory given by eqs.(86) and (87) are
related by a coordinate rotation.

Solution (III) For Region III For $\Delta >0$, $\varepsilon _{1}\geq
U>\varepsilon _{2}$, and $\varepsilon _{3},\bar{\varepsilon}_{3}$ complex.
Write the right-hand side of eq.(61) as $\beta ^{2}(\varepsilon
_{1}-U)(U-\varepsilon _{2})(U-\varepsilon _{3})(U-\bar{\varepsilon}_{3})$.

For Region IIIA bounded by $y_{3}$ [between $(1,0)$ and $(2,1/4)$] and $%
y_{2} $ [between $(2,1/4)$ and $(+\infty ,0)$], we let $\varepsilon
_{1}=x_{1},$ $\varepsilon _{2}=U_{1},$ $\varepsilon _{3}=x_{2},$ $\bar{%
\varepsilon}_{3}=\overline{x}_{2}$ from eqs.(73)-(75). Assuming the value of
$q$ to be given by $1/q=\varepsilon _{2}$ at $\phi =0$, we find the equation
for the trajectory to be

\begin{equation}
\frac{1}{q}=\frac{\varepsilon _{2}A(1+cn(2\gamma \phi ,k))+\varepsilon
_{1}B(1-cn(2\gamma \phi ,k))}{A(1+cn(2\gamma \phi ,k))+B(1-cn(2\gamma \phi
,k))},
\end{equation}

where

\begin{equation}
A^{2}=(\varepsilon _{1}-b_{1})^{2}+a_{1}^{2},
\end{equation}

\begin{equation}
B^{2}=(\varepsilon _{2}-b_{1})^{2}+a_{1}^{2},
\end{equation}

\begin{equation}
a_{1}^{2}=-\frac{(\varepsilon _{3}-\bar{\varepsilon}_{3})^{2}}{4},
\end{equation}

\begin{equation}
b_{1}=\frac{\varepsilon _{3}+\bar{\varepsilon}_{3}}{2},
\end{equation}

\begin{equation}
k^{2}=\frac{(\varepsilon _{1}-\varepsilon _{2})^{2}-(A-B)^{2}}{4AB},
\end{equation}

and

\begin{equation}
\gamma =\frac{1}{2}\beta \sqrt{AB}.
\end{equation}

For Region IIIB bounded by $y_{2}$ [between $(2,1/4)$ and $(+\infty ,0)$]
and $y_{3}$ [between $(2,1/4)$ and $(+\infty ,0)$], we let $\varepsilon
_{1}=U_{1},$ $\varepsilon _{2}=x_{1},$ $\varepsilon _{3}=x_{2},$ $\bar{%
\varepsilon}_{3}=\overline{x}_{2}$ from eqs.(73)-(75). We find the same
equation (88) for the trajectory, or

\begin{equation}
\frac{1}{q}=\frac{\varepsilon _{2}A(1-cn(2\gamma \phi ,k))+\varepsilon
_{1}B(1+cn(2\gamma \phi ,k))}{A(1-cn(2\gamma \phi ,k))+B(1+cn(2\gamma \phi
,k))},
\end{equation}

if we assume that the value of $q$ is given by $1/q=\varepsilon _{2}$ at $%
\phi =0$. Equation (95) is related to eq.(88) by a rotation of the
coordinates.

In all the Solutions for the light trajectories above, the Schwarzschild
limit is obtained if we let $\beta \rightarrow 0,$ $\varepsilon
_{1}\rightarrow \infty ,$ such that $\beta \sqrt{\varepsilon _{1}}%
\rightarrow 1$.

\section{Examples of Light Trajectories Including Special Cases}

A light beam is bent toward the black hole in Region I ($\Delta <0$) in the
R-N geometry in a manner similar to that in the Schwarzschild geometry, and
the total angle of deflection is given by eq.(84). An example of a light
beam bent by a charged black hole given by eq.(80) where $q$ is plotted
versus $\phi $ is shown in Fig.7 where the charged black hole is centered at
the origin. In this case $U_{1}=0.42922$ and $\beta ^{2}=0.1$ for which $%
q_{+}=0.8873$, $q_{-}=0.1127$, and the total angle of deflection $\Delta
\phi $ is $81.4^{\circ }$. Comparing this with the Schwarzschild case for
the same value of $U_{1}$ in which the black hole has no electric charge ($%
\beta ^{2}=0$) for which $\Delta \phi =90^{\circ }$ [3], we see that the
presence of electric charge in the black hole reduces the deflection angle
of the light beam.

On the upper boundary $y_{1}$ of Region I [between $(0,1/4)$ and $V$, see
Fig.6], $\varepsilon _{1}=\varepsilon _{2}$, $k^{2}=0$, the inverse distance
$U$ of the photon from the blackhole lies in the range $\varepsilon
_{1}=\varepsilon _{2}>\varepsilon _{3}\geq U>\varepsilon _{4}$, and the
trajectory is given by

\begin{equation}
\frac{1}{q}=\frac{\varepsilon _{3}(\varepsilon _{2}-\varepsilon
_{4})-\varepsilon _{2}(\varepsilon _{3}-\varepsilon _{4})\sin ^{2}(\gamma
\phi )}{(\varepsilon _{2}-\varepsilon _{4})-(\varepsilon _{3}-\varepsilon
_{4})\sin ^{2}(\gamma \phi )},
\end{equation}

where the roots of the quartic equation (60) are given by

\begin{equation}
\varepsilon _{1}=\varepsilon _{2}=\sqrt{\frac{-a}{3}}+\frac{1}{3}(\beta
^{-2}-U_{1}),
\end{equation}

\begin{eqnarray}
\varepsilon _{3} &=&U_{1},  \notag \\
\varepsilon _{4} &=&-2\sqrt{\frac{-a}{3}}+\frac{1}{3}(\beta ^{-2}-U_{1}),
\end{eqnarray}

where $\gamma $ and $a$ are given by eqs.(82) and (66). The trajectory of a
light beam is still one that is bent and the total angle of deflection $%
\Delta \phi $ is given by

\begin{equation}
\Delta \phi =-\pi +\frac{2}{\gamma }\sin ^{-1}\psi ,
\end{equation}

where%
\begin{equation}
\psi =\sqrt{\frac{(\varepsilon _{2}-\varepsilon _{4})\varepsilon _{3}}{%
(\varepsilon _{3}-\varepsilon _{4})\varepsilon _{2}}.}
\end{equation}

On the right boundary $y_{2}$ of Region I with Region II [between $(2/3,0)$
and $V$, see Fig.6], $\varepsilon _{2}=\varepsilon _{3}=U_{1}$, $k^{2}=1$,
the trajectories become asymptotic circles of radii $q_{u}$ given by $%
1/q_{u}=U_{1}$.

In Region II, the light trajectories given by eq.(86) or (87) are of the
wandering type an example of which is shown in Fig.8 for which $U_{1}=1$ and
$\beta ^{2}=0.1$, and for which $q_{+}=0.8873$ and $q_{-}=0.1127$. It is
seen that starting from $q_{0}=1$, the trajectory can repeatedly enter and
emerge from the region in which $q$ is between $q_{+}$ and $q_{-}$. The
physical interpretation of this behavior is not clear [1,2]. For the same
value of $U_{1}$ but with $\beta ^{2}=0$, the trajectory would simply be one
that terminates at the black hole.

On the boundary $y_{2}$ from $V$ to $(+\infty ,0)$ [see Fig.6] between
Regions IIA and IIB, and between Regions IIIA and IIIB, $\varepsilon
_{1}=\varepsilon _{2}=U_{1}$, $k^{2}=0$, and the trajectories are given by
circles of radii $q_{c}$ given by $1/q_{c}=U_{1}$.

On the upper boundary $y_{1}$ of Region IIB [between $V$ and $(+\infty ,0)$,
see Fig.6],

\begin{eqnarray*}
\varepsilon _{1} &=&U_{1}, \\
\varepsilon _{2} &=&\varepsilon _{3}=\sqrt{\frac{-a}{3}}+\frac{1}{3}(\beta
^{-2}-U_{1}), \\
\varepsilon _{4} &=&-2\sqrt{\frac{-a}{3}}+\frac{1}{3}(\beta ^{-2}-U_{1}),
\end{eqnarray*}

where $a$ is given by eq.(66), from which $k^{2}=1$ from eq.(81). From
eq.(87), where $\gamma $ is given by eq.(82), the trajectories become
asymptotic circles of radii $q_{u}$ given by $1/q_{u}=\varepsilon _{2}$ as $%
\gamma \phi \rightarrow \infty $ since $sn(\gamma \phi ,1)=\tanh (\gamma
\phi )$.

On the boundary $y_{3}$ between Region IIA and IIIA [between $(1,0)$ and $%
(2,1/4)$, see Fig.6], $\varepsilon _{3}=\varepsilon _{4}=0$, $k^{2}=0$, and
the trajectories are given by

\begin{equation}
\frac{1}{q}=\frac{\varepsilon _{1}\varepsilon _{2}}{\varepsilon
_{1}-(\varepsilon _{1}-\varepsilon _{2})\sin ^{2}(\gamma \phi )},
\end{equation}

where $\gamma $ is given by eq.(82), and

\begin{eqnarray}
\varepsilon _{1} &=&\beta ^{-2}-U_{1},  \notag \\
\varepsilon _{2} &=&U_{1}.
\end{eqnarray}

On the boundary $y_{3}$ between Region IIB and IIIB [between $(2,1/4)$ and $%
(+\infty ,0)$, see Fig.6], $\varepsilon _{3}=\varepsilon _{4}=0$, $k^{2}=0$,
the trajectories are given by

\begin{equation}
\frac{1}{q}=\frac{\varepsilon _{1}\varepsilon _{2}}{\varepsilon
_{2}+(\varepsilon _{1}-\varepsilon _{2})\sin ^{2}(\gamma \phi )},
\end{equation}

where $\gamma $ is given by eq.(82), and

\begin{eqnarray}
\varepsilon _{1} &=&U_{1},  \notag \\
\varepsilon _{2} &=&\beta ^{-2}-U_{1}.
\end{eqnarray}

The trajectories given by eqs.(101) and (103) are closed curves. An example
of a light trajectory given by eq.(101) is shown in Fig.9 for which $%
U_{1}=1.2$, $\beta ^{2}=0.13889$, $q_{+}=0.83333$, $q_{-}=0.16667$. The
closed curve has $q_{\max }=q_{+}$ and $q_{\min }=q_{-}$. At the
intersection point $(2,1/4)$ of $y_{3}$ and $y_{2}$ on which $\varepsilon
_{1}=\varepsilon _{2}=2$, we have a circular orbit of radius $q_{c}=1/2$ or $%
r_{c}=GM/c^{2}$.

At the vertex point $V=(4/3,9/32)$, $\varepsilon _{1}=\varepsilon
_{2}=\varepsilon _{3}=U_{1}=4/3$, $\varepsilon _{4}=-4/9$, $\beta
^{2}=9/32=0.28125$, $\gamma =0$ from eq.(82), and eqs.(80), (86) and (87)
all become a circular orbit with a radius $q_{c}=1/U_{1}=3/4$ or $%
r_{c}=3GM/(2c^{2})$.

\section{Summary}

We have characterized all trajectories of an electrically neutral particle
(planet) around an electrically charged black hole in the Reissner-Nordstr%
\"{o}m geometry by three dimensionless parameters $e^{2}$, $s^{2}$, and $%
\beta ^{2}$ defined by eqs.(1), (2) and (6), and placed them on a parameter
space $(e^{2},s^{2})$ shown in Fig.1 that consists of two regions, called
Regions I and II, where analytic solutions for the trajectories given by
eqs.(33) and (40) for Region I, and by eq.(41) for Region II, apply.
Analytic expressions for the boundaries of Regions I and II for two specific
values of the dimensionless charge $\beta ^{2}=0$ and $1/4$ are given by
eqs.(28) and (30), and the corresponding curves for the boundaries are shown
in Fig.1. Examples of these particle trajectories in Regions I and II are
shown in Figs.2-5. Of particular interest are the following results on the
effect of the presence of a net electric charge ($\beta ^{2}>0$) on the
black hole compared to the case when the black hole is electrically neutral (%
$\beta ^{2}=0$ or the Schwarzschild case): (1) the precession angle of an
elliptic-type orbit decreases, (2) the deflection angle of a hyperbolic-type
orbit also decreases, and (3) a wandering-type orbit replaces a terminating
orbit.

We have characterized the light trajectories near a charged black hole by
two dimensionless parameters $U_{1}$ and $\beta ^{2}$, where $U_{1}$ is
defined just prior to eq.(61). The parameter space $(U_{1},\beta ^{2})$ is
shown in Fig.6 and is divided into three principal Regions, called I, II and
III, with Regions II and III divided into sub-regions A and B. Equations
(80), (86), (87), (88), and (95) apply to Regions I, IIA, IIB, IIIA, and
IIIB respectively. Figures 7, 8 and 9 show examples of a light trajectory
near a charged black hole in Regions I, II and on the boundary of Regions II
and III. The boundary curves separating the various regions in Fig.6 are
given by eqs.(76), (78) and (79). Of particular interest are the following
results on the effect of the presence of a net electric charge on the black
hole compared to the case when the black hole is electrically neutral: (1)
the bending of a light ray decreases, (2) a wandering-type trajectory
replaces a terminating one, and (3) there are closed trajectories that are
not circular.

Acknowledgements

I am very grateful to Drs. Dave Kuebel and Clark Carroll for many valuable
comments, corrections and suggestions, and to Dr. Krsna Dev for valuable
comments and technical help. I would also like to thank Drs. C. L\"{a}%
mmerzahl, L. Iorio and M. Azreg-A\"{\i}nou for drawing my attention to their
work after the appearance of this paper in arXiv:1402.1756v1 [gr-qc] (2014).

\bigskip

\bigskip

\bigskip

References

*Electronic address: fhioe@sjfc.edu

[1] M.P. Hobson, G. Efstathiou and A.N. Lasenby: General Relativity,
Cambridge University Press, 2006, Chapters 12.

[2] S. Chandrasekhar: The Mathematical Theory of Black Holes, Oxford
University Press, 1992, Chapter 5.

[3] F.T. Hioe and D. Kuebel, Phys. Rev. D 81, 084017 (2010).

[4] F.T. Hioe and D. Kuebel, arXiv:1207.7041v1 (2012).

[5] P.F. Byrd and M.D. Friedman: Handbook of Elliptic Integrals for
Engineers and Scientists, 2nd Edition, Springer-Verlag, New York, 1971.

[6] Y. Hagihara, J. Astron. Geophys. 8, 67-176 (1931).

[7] E.T. Whittaker: A Treatise on the Analytical Dynamics of Particles and
Rigid Bodies, 4th Edition, Dover, New York 1944, Chapter XV.

[8] S. Grunau and V. Kagramanova, Phys. Rev. D 83, 044009 (2011).

[9] E. Hackmann, V. Kagramanova, J. Kunz and C. L\"{a}mmerzahl, Phys. Rev. D
78, 124018 (2008).

[10] D. Pugliese, H. Quevedo and R. Ruffini, Phys. Rev. D 83, 104052 (2011).

[11] L. Iorio, Gen. Relativ. Gravit. 44, 1753 (2012).

[12] G.W. Gibbons and M. Vyska, Class. Quantum Grav. 29, 065016 (2012).

[13] M. Azreg-A\"{\i}nou, Phys. Rev. D 87, 024012 (2013).

[14] A.R. Forsyth, Proc. Roy. Soc. Lond. A 97, 145 (1920).

[15] Ref.3 studied the cases for $0\leq e^{2}\leq \infty $ only. See ref.4
for the more complete results for $-\infty \leq e^{2}\leq \infty $.

[16] C. Darwin, Proc. Roy. Soc. Lond. A249, 180 (1958), ibid. A263, 39
(1961).

[17] F.T. Hioe and D. Kuebel, arXiv:1208.0260v1 (2012).

[18] F.T. Hioe, Phys. Lett. A 373, 1506 (2009).

[19] J.L. Martin: General Relativity, Revised Edition, Prentice Hall, New
York 1996, Chapter 4.

[20] F.T. Hioe and D. Kuebel, arXiv:1008.1964v1 (2010).

[21] C.E. Carroll obtained this result independently, private communication.

\begin{figure}[p]
\includegraphics{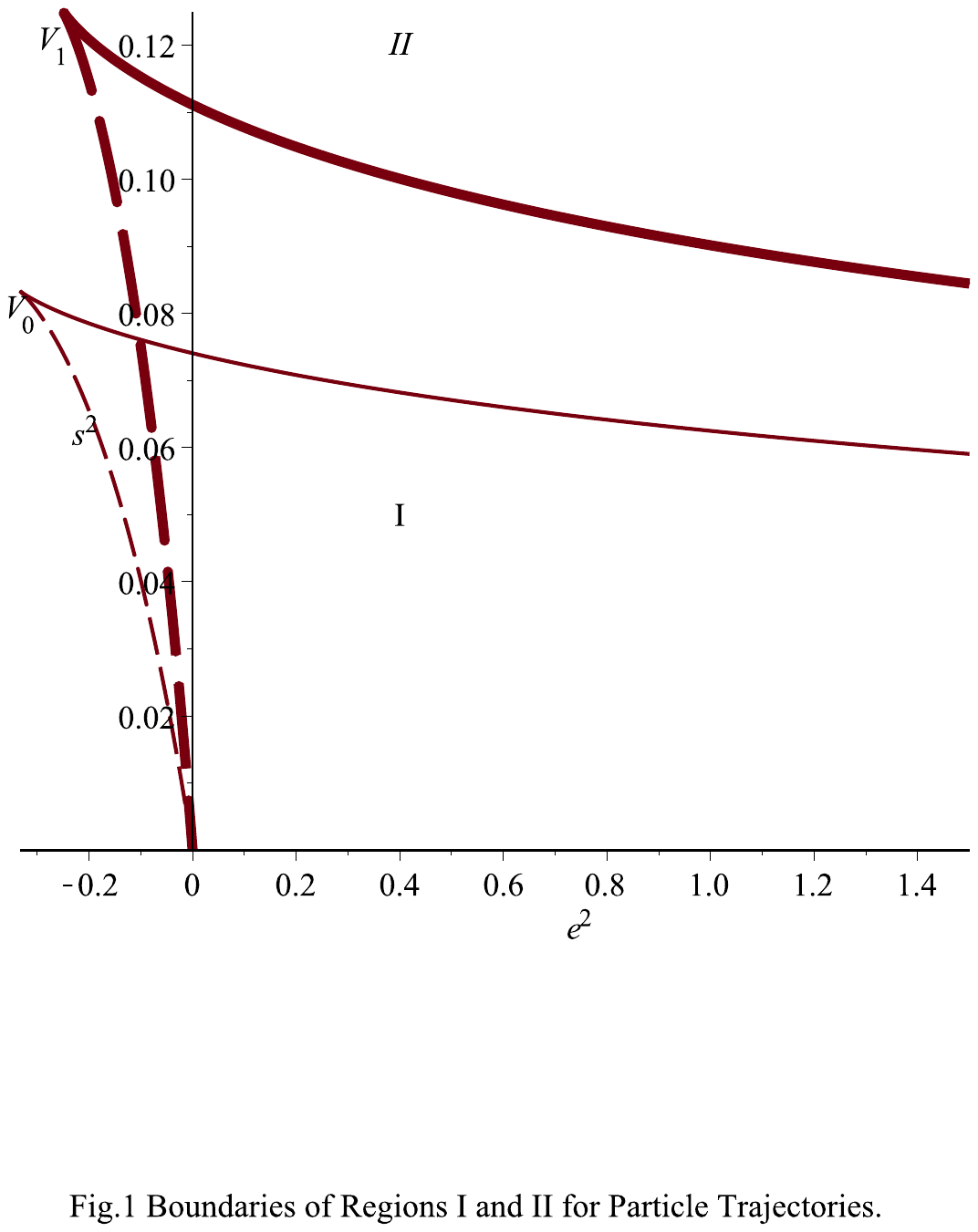}
\end{figure}

\begin{figure}[p]
\includegraphics{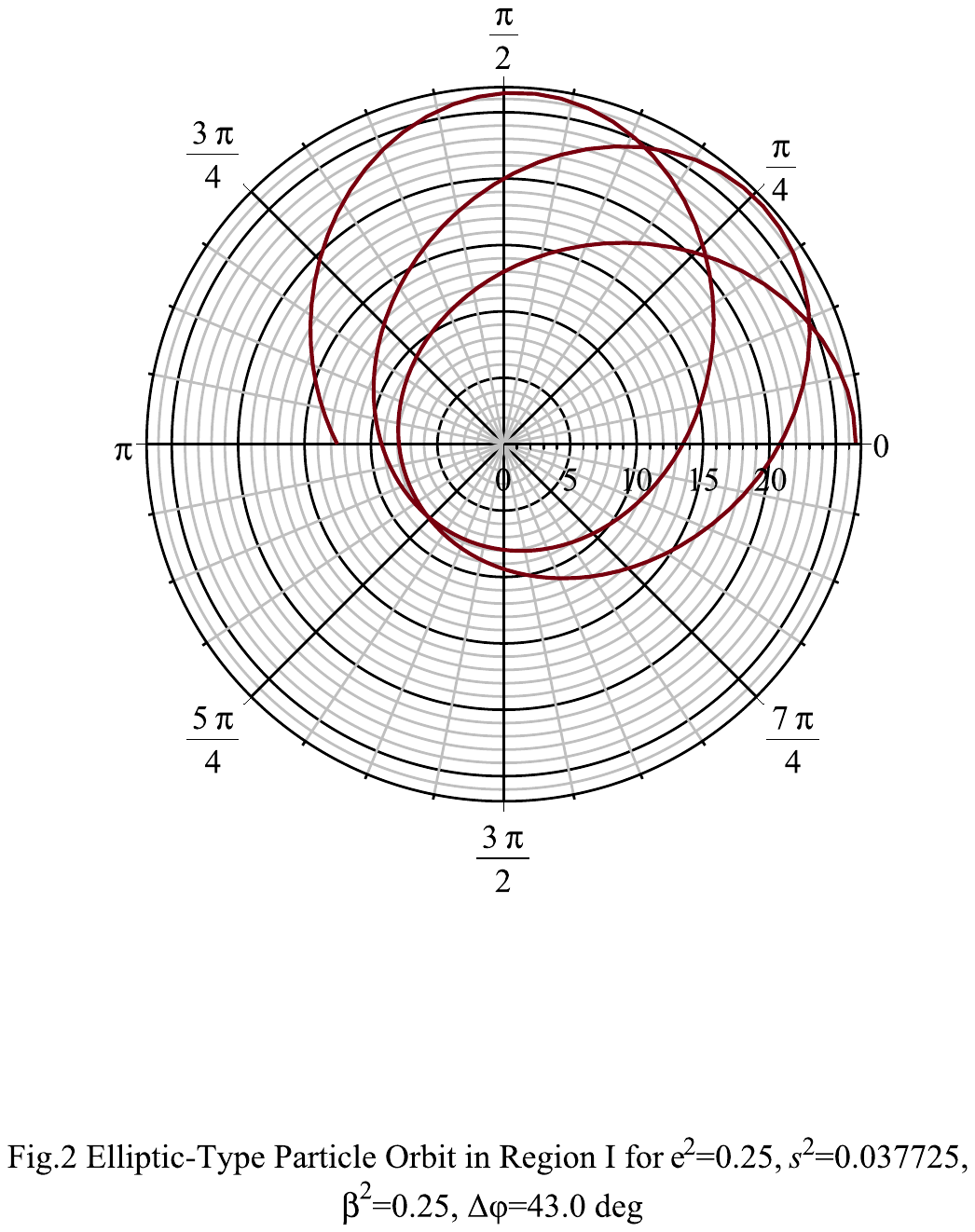}
\end{figure}

\begin{figure}[p]
\includegraphics{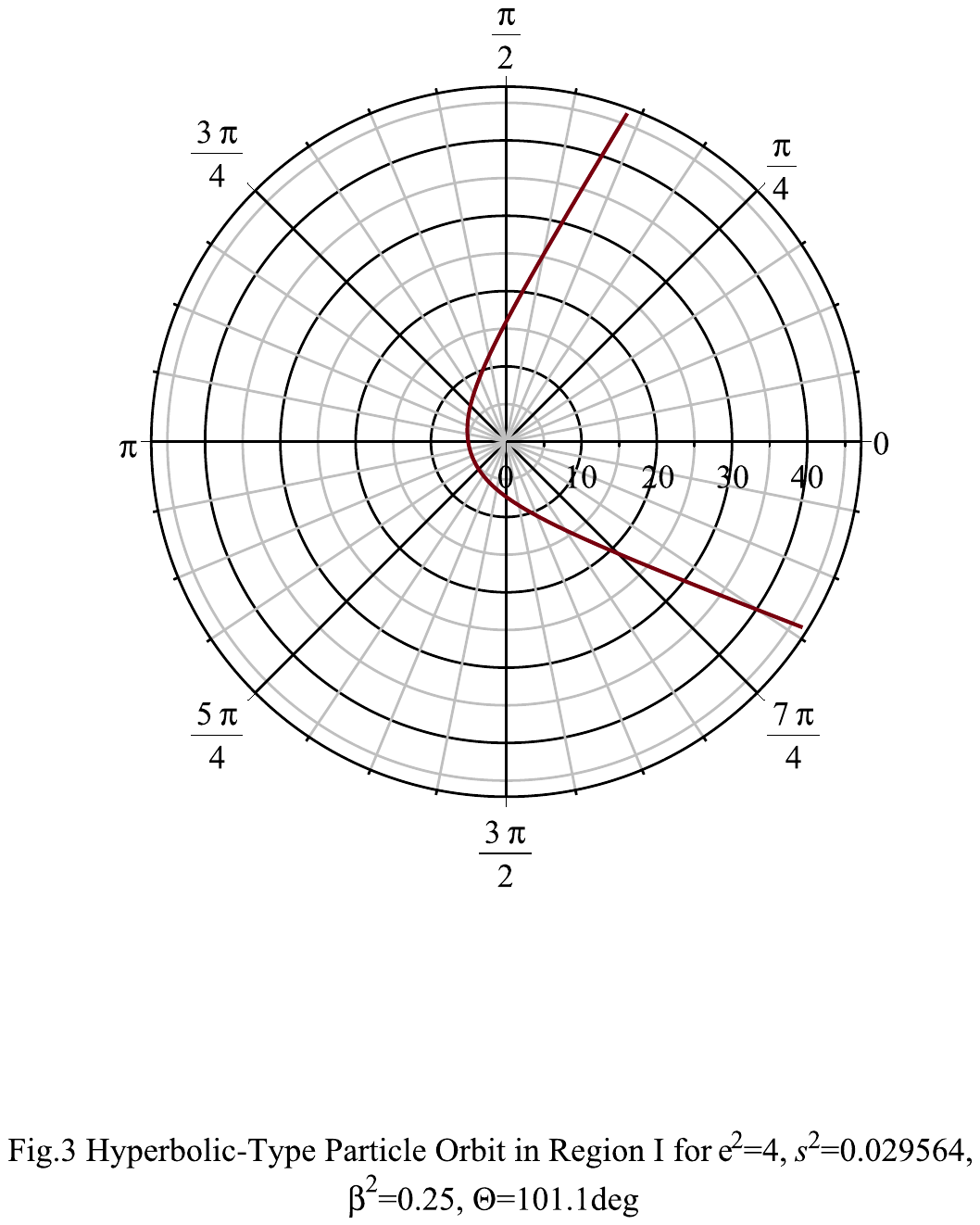}
\end{figure}

\begin{figure}[p]
\includegraphics{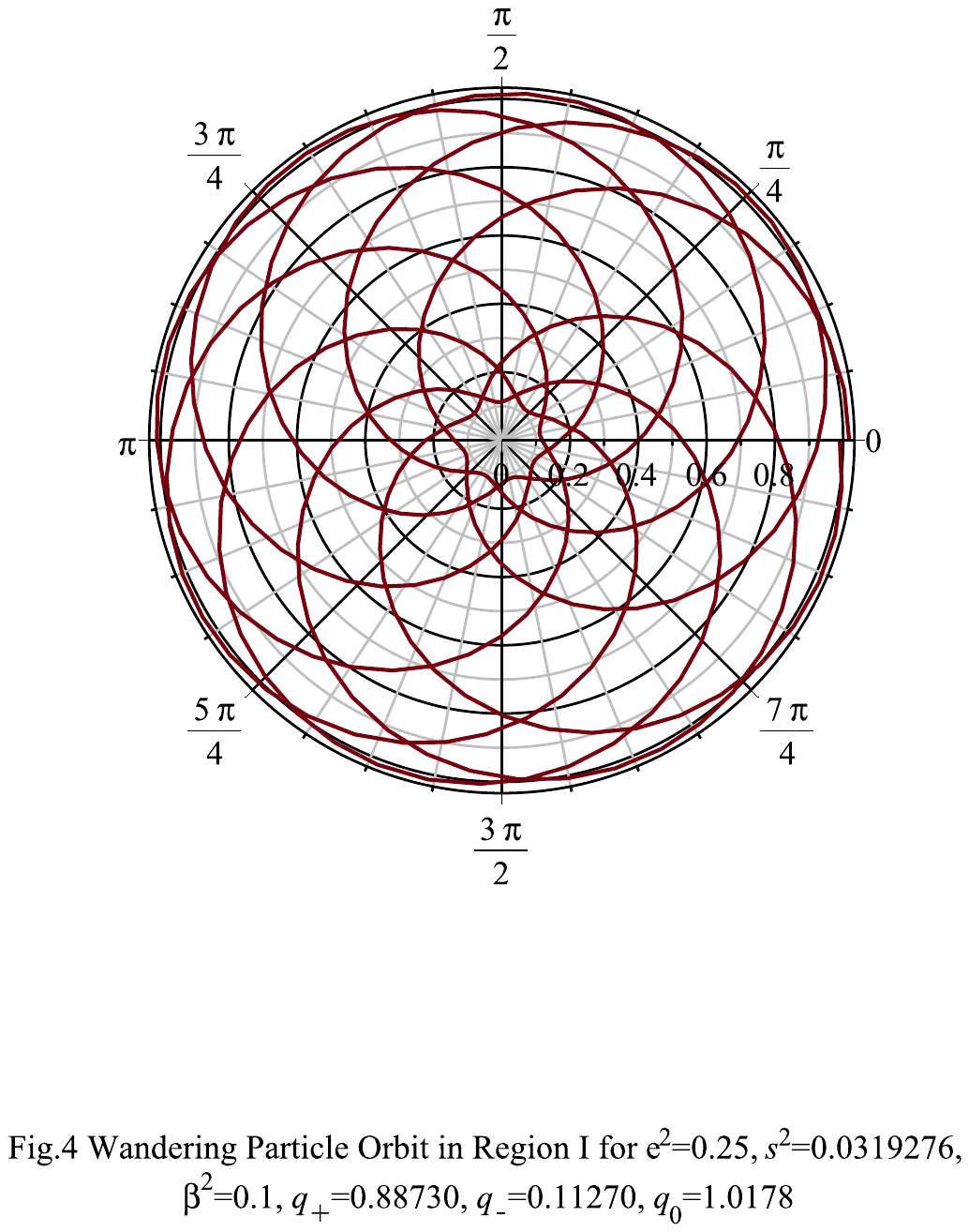}
\end{figure}

\begin{figure}[p]
\includegraphics{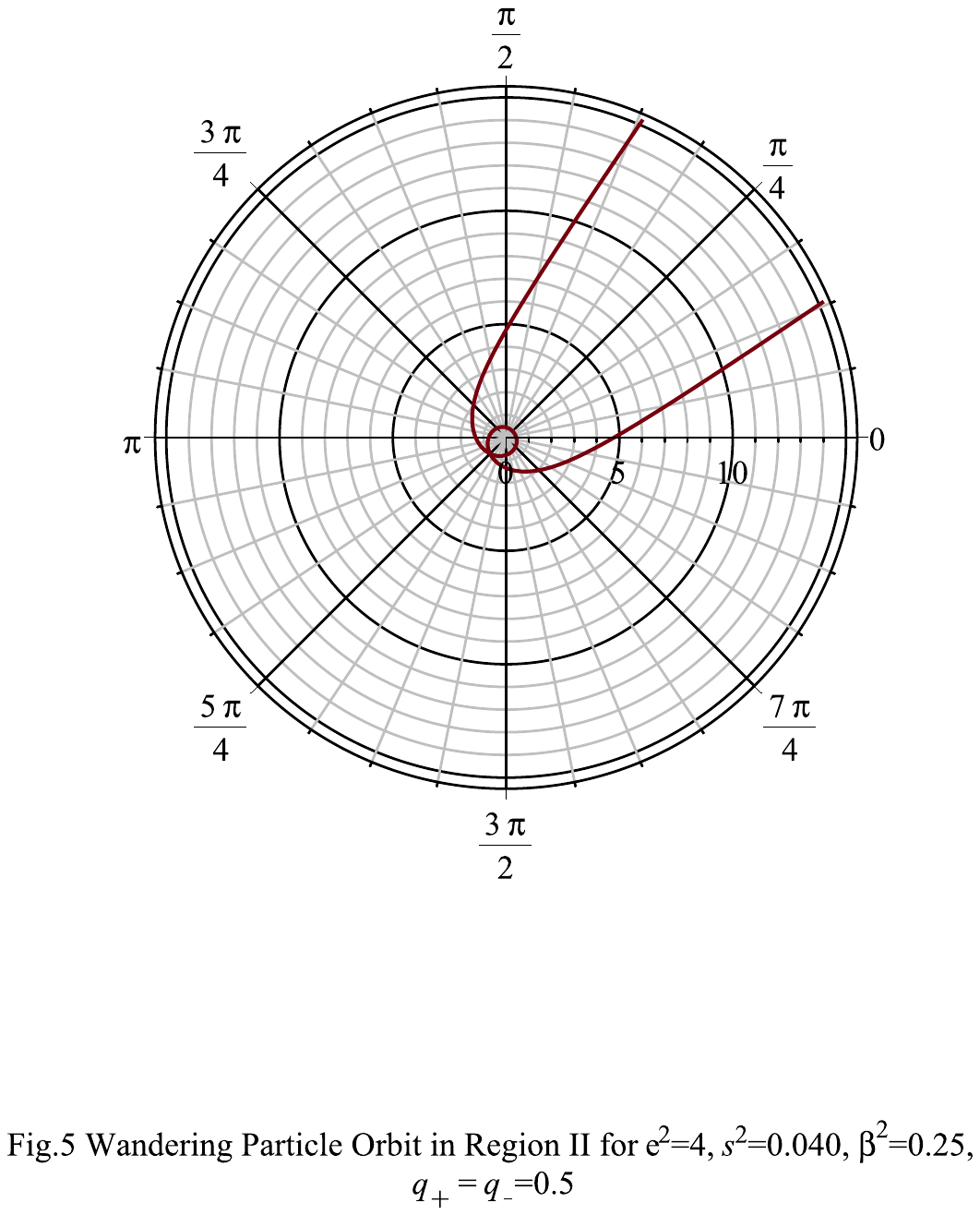}
\end{figure}

\begin{figure}[p]
\includegraphics{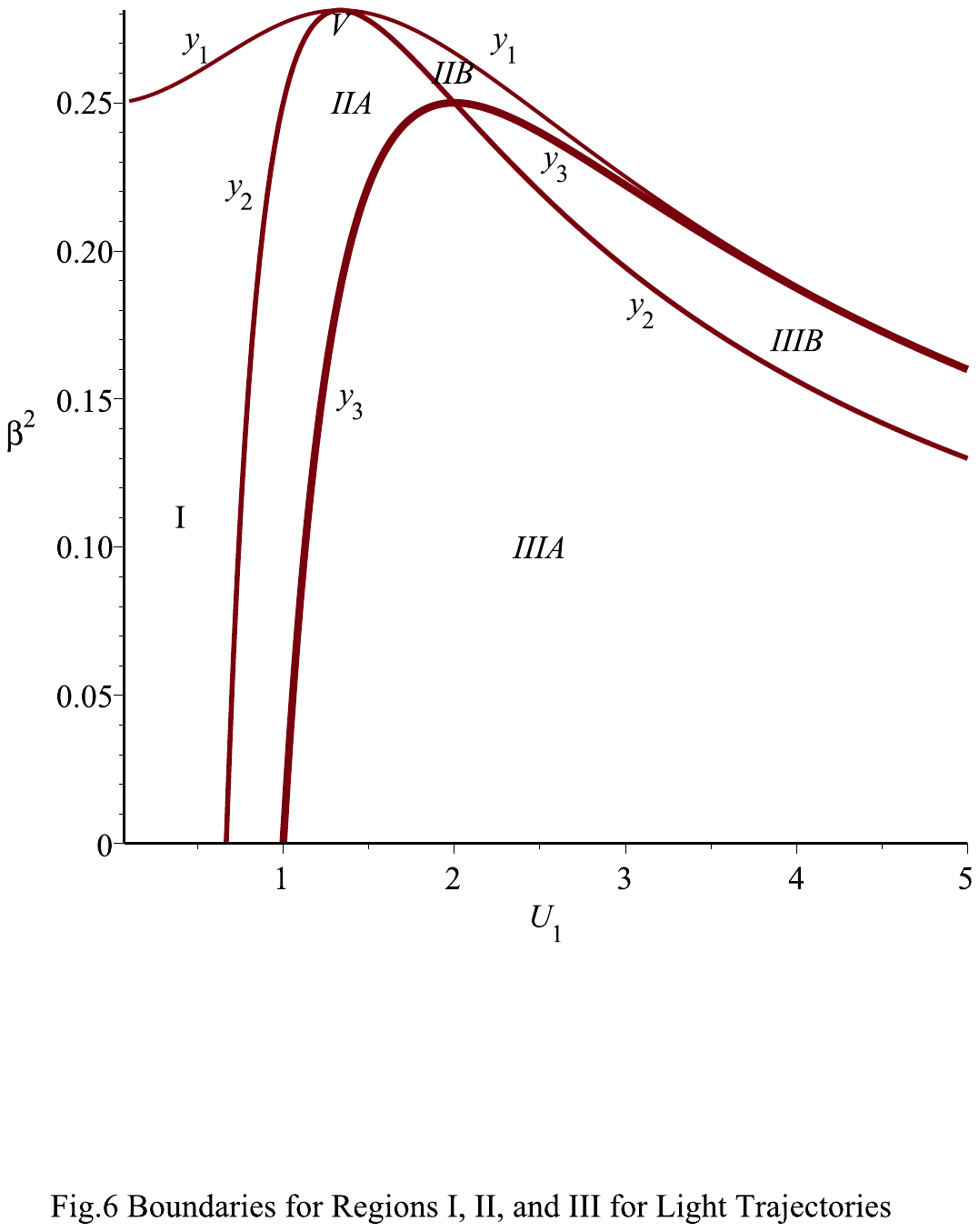}
\end{figure}

\begin{figure}[p]
\includegraphics{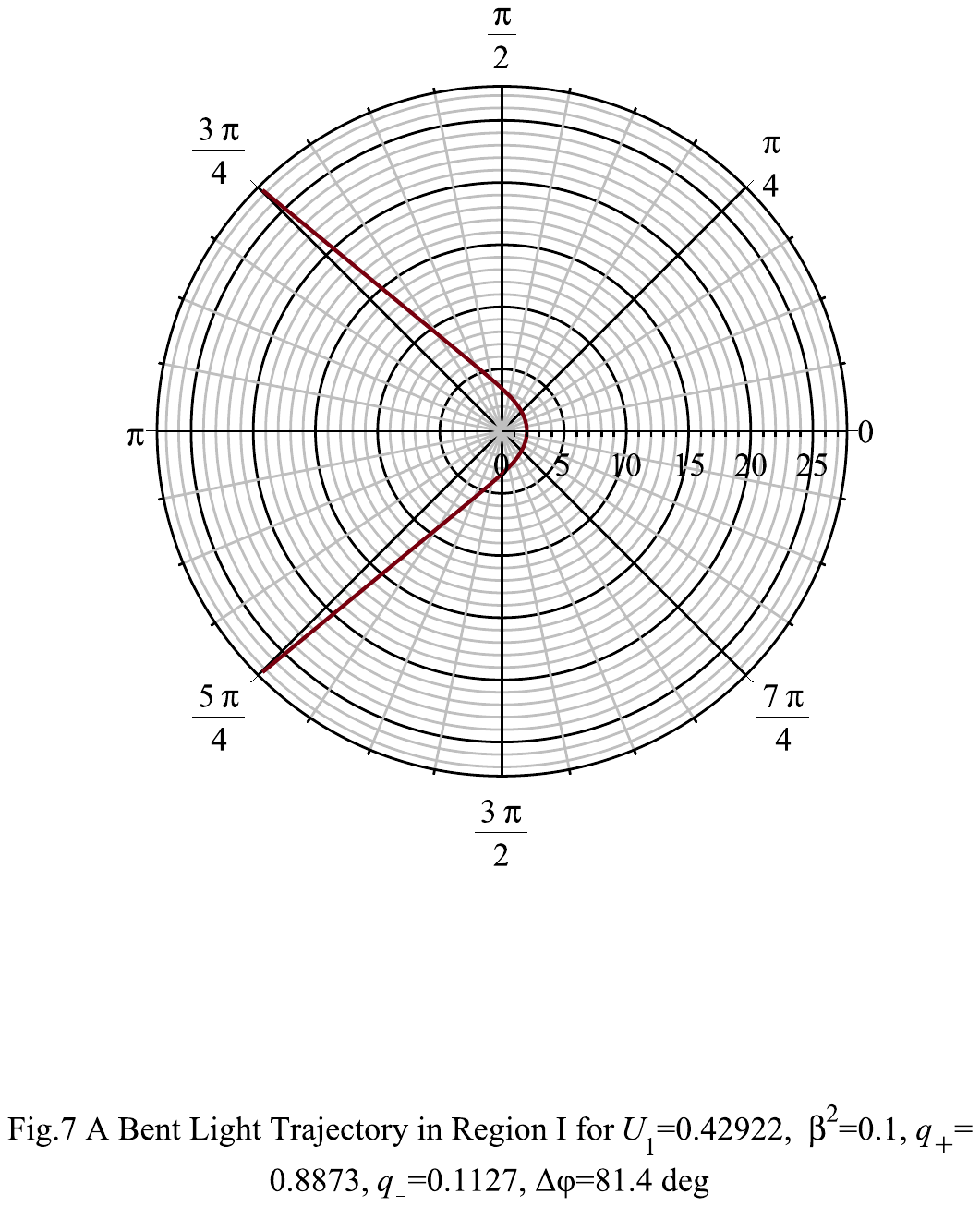}
\end{figure}

\begin{figure}[p]
\includegraphics{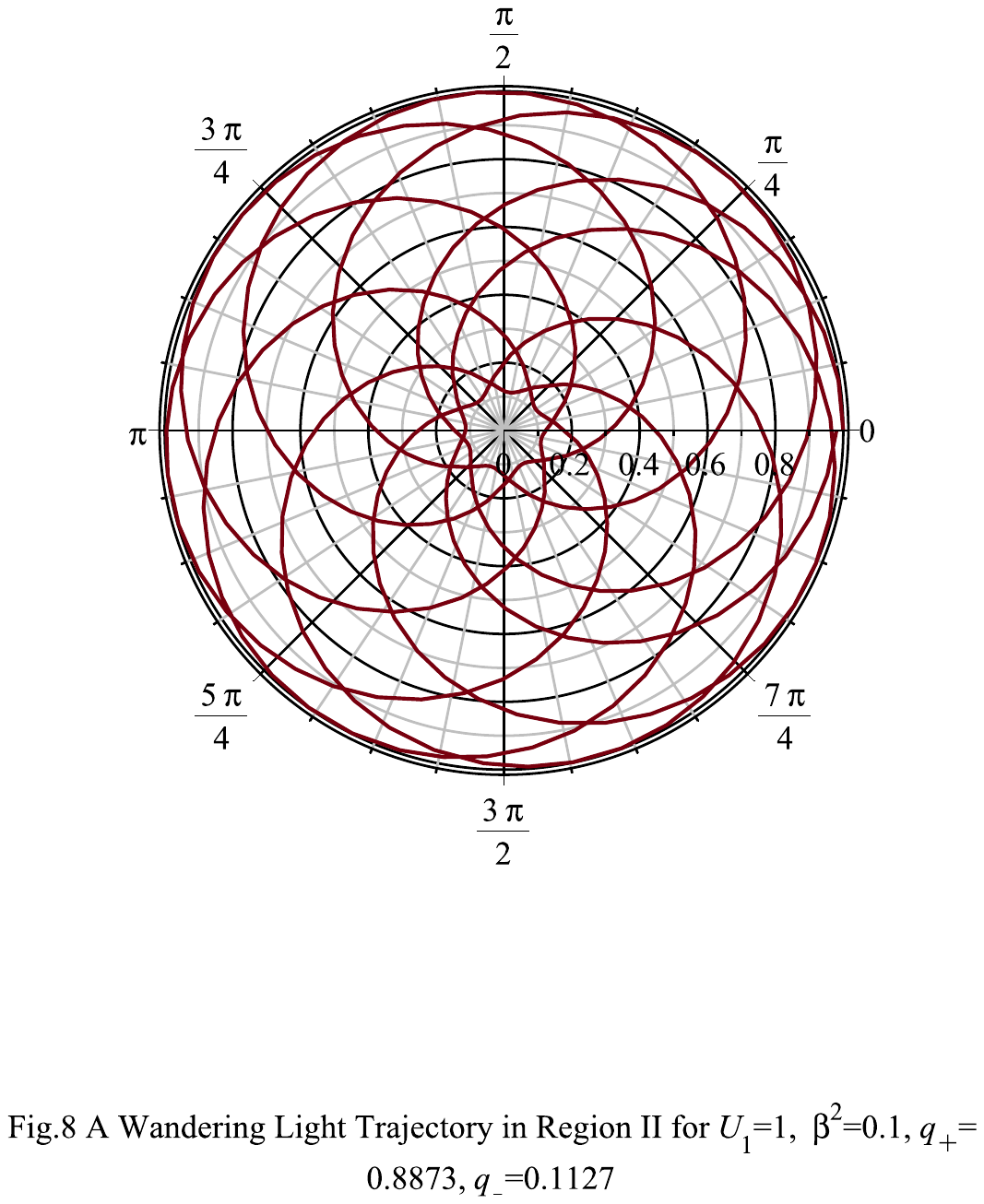}
\end{figure}

\begin{figure}[p]
\includegraphics{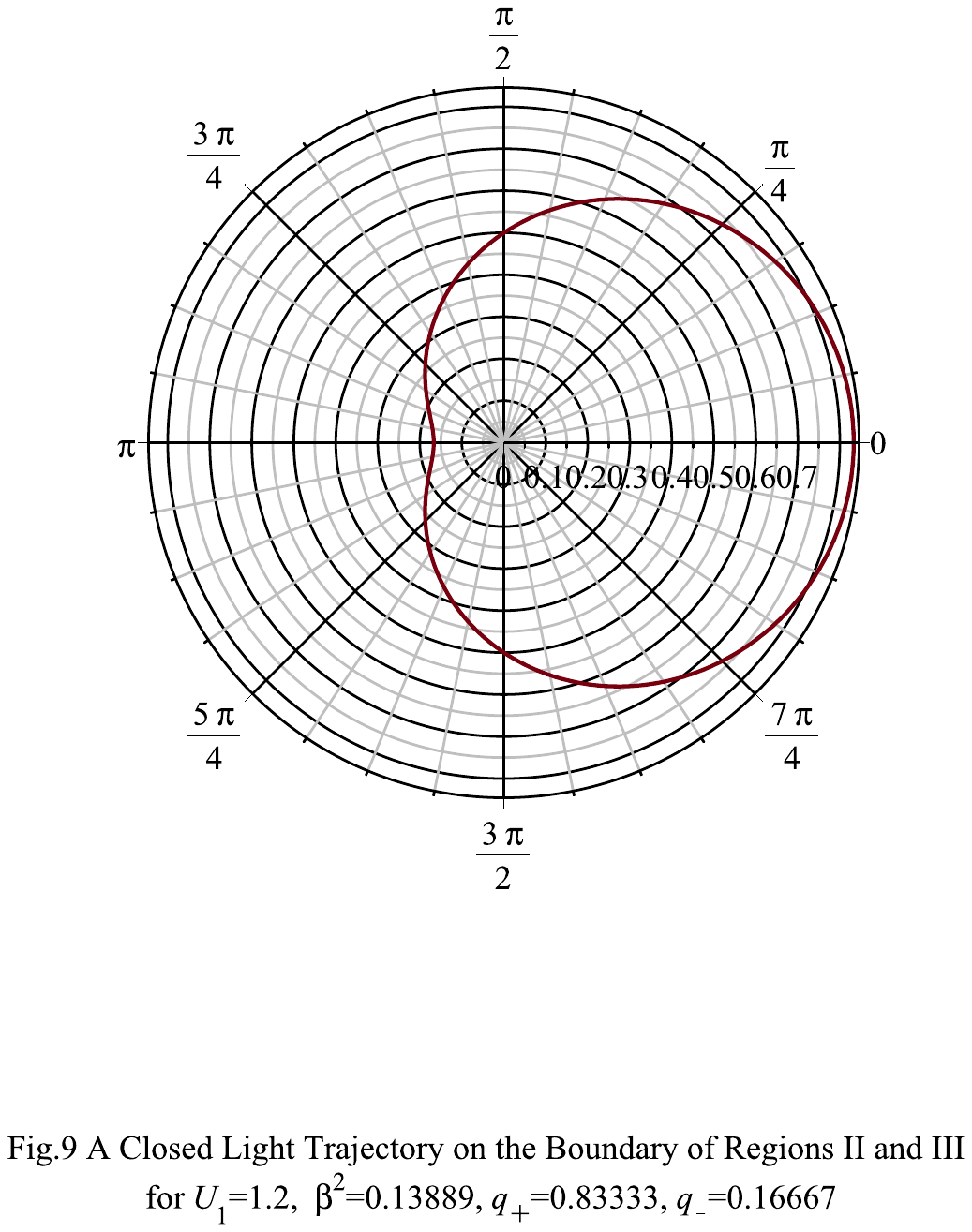}
\end{figure}

\end{document}